\documentclass{aa}  

\usepackage{graphicx}
\usepackage{txfonts}
\usepackage{natbib}
\usepackage{hyperref}

\usepackage{threeparttable}
\usepackage{supertabular}

\usepackage{amsmath}
\usepackage{subcaption} 
\usepackage{upgreek}
\usepackage{float}

\usepackage{tablefootnote}
\usepackage{tikz}

\newcommand{\cref}{\ref}

\usepackage[normalem]{ulem}

\begin{document}


\title{Temporal and spectral variations of the X-ray pulsar Cen X-3 observed by NuSTAR}

\author{Qi Liu\inst{1,2}
\and Wei Wang\inst{2}
\and Andrea Santangelo\inst{1}
\and Lingda Kong\inst{1}
\and Long Ji\inst{3}
\and Lorenzo Ducci\inst{1}
}
\institute{Institut für Astronomie und Astrophysik, Universität Tübingen, Sand 1, 72076 Tübingen, Germany \\
\email{qi.liu@mnf.uni-tuebingen.de}
\and
Department of Astronomy, School of Physics and Technology, Wuhan University, Wuhan 430072, China \\ \email{wangwei2017@whu.edu.cn}
\and 
School of Physics and Astronomy, Sun Yat-Sen University, Zhuhai 519082, China
}
\date{ }

\abstract
{We report a time-resolved analysis of the accreting X-ray pulsar Cen X-3 using observations carried out by NuSTAR, which cover approximately two binary orbits in different intensity states. The pulse profile is relatively stable over the orbital phase and shows energy dependence. It has an obvious double-peaked shape in the energy band below 15 keV ---with the second pulse peak decreasing as energy increases--- and is gradually dominated by a single peak in higher energy bands. We find that the pulse profile in the energy band of 3-5 keV at high-intensity states shows a subtle triple-peaked shape, with the main peak divided into two subpeaks. We also find a positive correlation between the pulse fraction and both energy and flux. Our spectral analysis reveals that the spectra can be well described by the continuum of Fermi-Dirac cutoff and NPEX models, and the cyclotron line is detected with the centroid energies varying from 26 keV to 29 keV, along with the iron emission line around 6.4 keV. We investigated the dependence between the cyclotron resonant scattering feature (CRSF) centroid energy and luminosity and discuss the theoretical critical luminosity. Although the variation of $E_{\rm cyc}- L_X$ is not distinct, there is a possibility that the critical luminosity lies within the range of $\sim (0.5-4)\times 10^{37}$ erg s$^{-1}$ in the band of $4-78$ keV. The photon index shows a strong positive correlation with luminosity. Our orbital-phase analysis reveals that the spectral parameters show orbital variability, and the highly variable photoelectric absorption may indicate the existence of clumpy stellar winds. 
}


\maketitle

\section{Introduction}
\label{sec:introduction}

Cen X-3, a persistently accreting X-ray pulsar, was first identified by the Uhuru X-ray Explorer Satellite \citep{1971ApJ...167L..67G,1972ApJ...172L..79S}. Cen X-3 is an eclipsing high-mass X-ray binary (HMXB) with a spin period of approximately 4.8 seconds and an orbital period of around 2.1 days (also see \citealt{2023A&A...675A.135K,2023JHEAp..38...32L}). The system exhibits a decay rate of $\sim 1.799 \times 10^{-6}$ yr\(^{-1}\) \citep{2015AA...577A.130F}. The binary system consists of a neutron star with a mass of approximately \(1.21 \pm 0.21 \ M_{\odot}\) and an optical companion O6-8 III supergiant star \citep{1974ApJ...192L.135K,1979ApJ...229.1079H} with a mass of about \(20.5 \pm 0.7 \ M_{\odot}\) and a radius of 12 \(R_{\odot}\) \citep{1999MNRAS.307..357A}. The estimated distance to the binary system is approximately 6.9 kpc based on Gaia measurements \citep{2021MNRAS.507.3899V,2021A&A...649A...1G}. The high X-ray luminosity (\(\sim5 \times 10^{37}\) erg s\(^{-1}\), \citealt{2008ApJ...675.1487S}), a secular spin-up trend in this system, and the quasi-periodic oscillations at 40 mHz \citep{2008ApJ...685.1109R,1991PASJ...43L..43T,2022MNRAS.516.5579L} may suggest the existence of an accretion disk.

Cyclotron resonant scattering features (CRSFs, \citealt{2019A&A...622A..61S}), also called cyclotron lines, are identified as absorption features in the continuum spectrum of X-ray pulsars. In highly magnetized accreting neutron stars, electrons perpendicular to the strong magnetic fields are quantized into discrete Landau levels. These absorption-like features form through the resonant scattering of photons on these electrons in the line-forming region. Measurements of the energy of the cyclotron line provide a direct method for determining the magnetic strength of neutron stars, because the cyclotron line energy is linked to the magnetic field, which can be expressed as follows:
\begin{equation}
E_{\mathrm{cyc}} \approx \frac{11.57 \ \mathrm{keV}}{(1+z)} B_{12},
\end{equation}
where \(B_{12}\) represents the magnetic field strength of the neutron star in the scattering region in units of \(10^{12}\) G, and \(z\) denotes the gravitational redshift. \cite{1998A&A...340L..55S} estimated the magnetic field of the neutron star surface for Cen X-3, finding it to be in the range of (2.4-3.0) $\times 10^{12}$ G using \(E_{\mathrm{cyc}} \sim\) 28 keV.

The X-ray spectra in Cen X-3 are typically modeled by a power law with a high-energy cut-off, an iron emission line, and a soft excess below 1 keV \citep{1983ApJ...270..711W,2000ApJ...530..429B,2008ApJ...675.1487S}. An absorption line feature around 30 keV in the continuum spectra was first observed by \cite{1992ApJ...396..147N} using Ginga observations. The presence of the CRSF was later confirmed by \cite{1998A&A...340L..55S} through BeppoSAX satellite observations. Subsequently, the dependence of CRSF parameters on pulse phase in Cen X-3 was explored by \cite{2000ApJ...530..429B} with BeppoSAX and by \cite{2008ApJ...675.1487S} with Rossi X-ray Timing Explorer (RXTE). New measurements of the cyclotron line energy were also provided by \cite{2021MNRAS.500.3454T} using Suzaku and NuSTAR.  \cite{Tamba_2023} also presented a detailed time-resolved analysis with NuSTAR observation data from 2015. In the present paper, we further study the timing and spectral properties of Cen X-3 using new NuSTAR observations from 2022, which cover about two orbital periods.

The paper is organized as follows: In Section 2, we briefly describe the NuSTAR observations and data reduction. We show the temporal analysis in Section 3.  We present our analysis of the spectral fitting in Section 4, showing the evident cyclotron lines. In Section 5, we focus on the dependence of CRSF energies on luminosity. In addition, the orbital dependence of CRSF and continuum parameters are discussed. We summarize our conclusions in Section 6. 

\section{Observations and data reduction} \label{sec:Data}

\begin{figure}
    \centering
    \includegraphics[width=.5\textwidth]{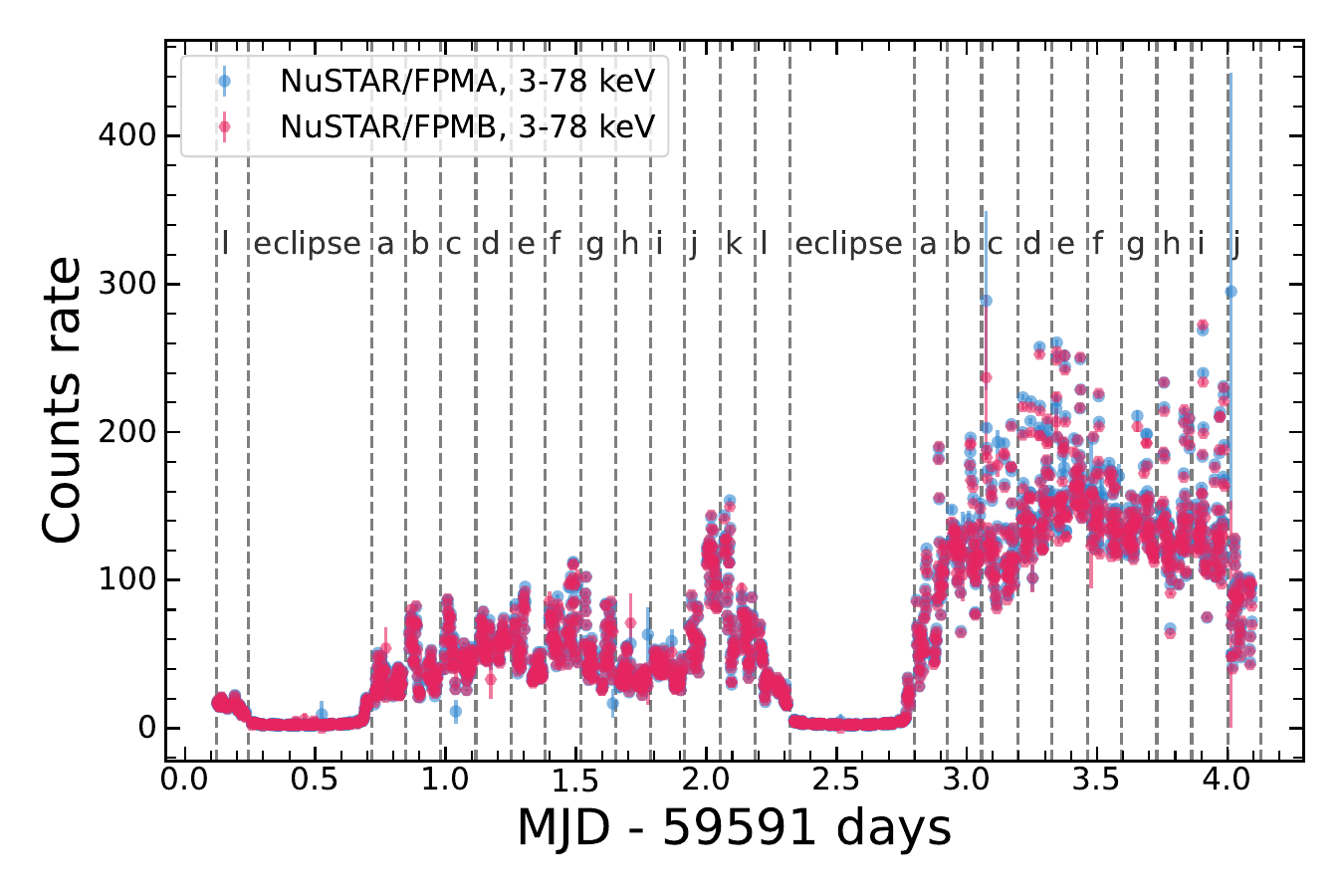}
    \caption{ FPMA and FPMB count rates at 3-78 keV of NuSTAR in 2022 with a time resolution of 100s.}.
    \label{fig:counts}
\end{figure}

The NuSTAR  observations were carried out by \citet{2013ApJ...770..103H} from January 12 to January 16, 2022, with a total exposure time of 189 ks (ObsID: 30701019002). We processed the data with the standard NuSTAR analysis software in HEASoft 6.32.1. The source regions were extracted as circular regions with a radius of 180" from the source center, and the background regions were extracted with an identical area in the off-source regions. We then applied the Solar System barycentric correction to the cleaned event files.

Light curves were extracted with a bin size of 10 ms, and spectra were generated from the barycentric corrected event files using the task \textit{nuproducts}. Figure \cref{fig:counts} shows the counts rate in 3-78 keV, which cover about two binary orbits in different intensity states. Here, we excluded the eclipse data and divided the observation into 23 different intervals based on the duration of $\sim$10 ks in preparation for subsequent analysis; these intervals are shown in Figure \cref{fig:counts}. Before the temporal analysis, the shift in arrival time due to binary orbital motion was corrected for each light curve using the orbital parameters from Fermi/GBM\footnote{\url{https://gammaray.msfc.nasa.gov/gbm/science/pulsars/lightcurves/cenx3.html}, Orbital parameters from M.H.Finger (HEAD 2010 poster).}. All of the spectral analyses in this paper were performed with XSPEC 12.13.1\citep{1996ASPC..101...17A}. The uncertainties are calculated as 90\% confidence levels using the Markov chain Monte Carlo (MCMC) method with ten walkers and a chain length of 10 000.

\section{Temporal analysis}\label{sec:timing}

\begin{figure*}
    \centering
    \includegraphics[width=.33\textwidth]{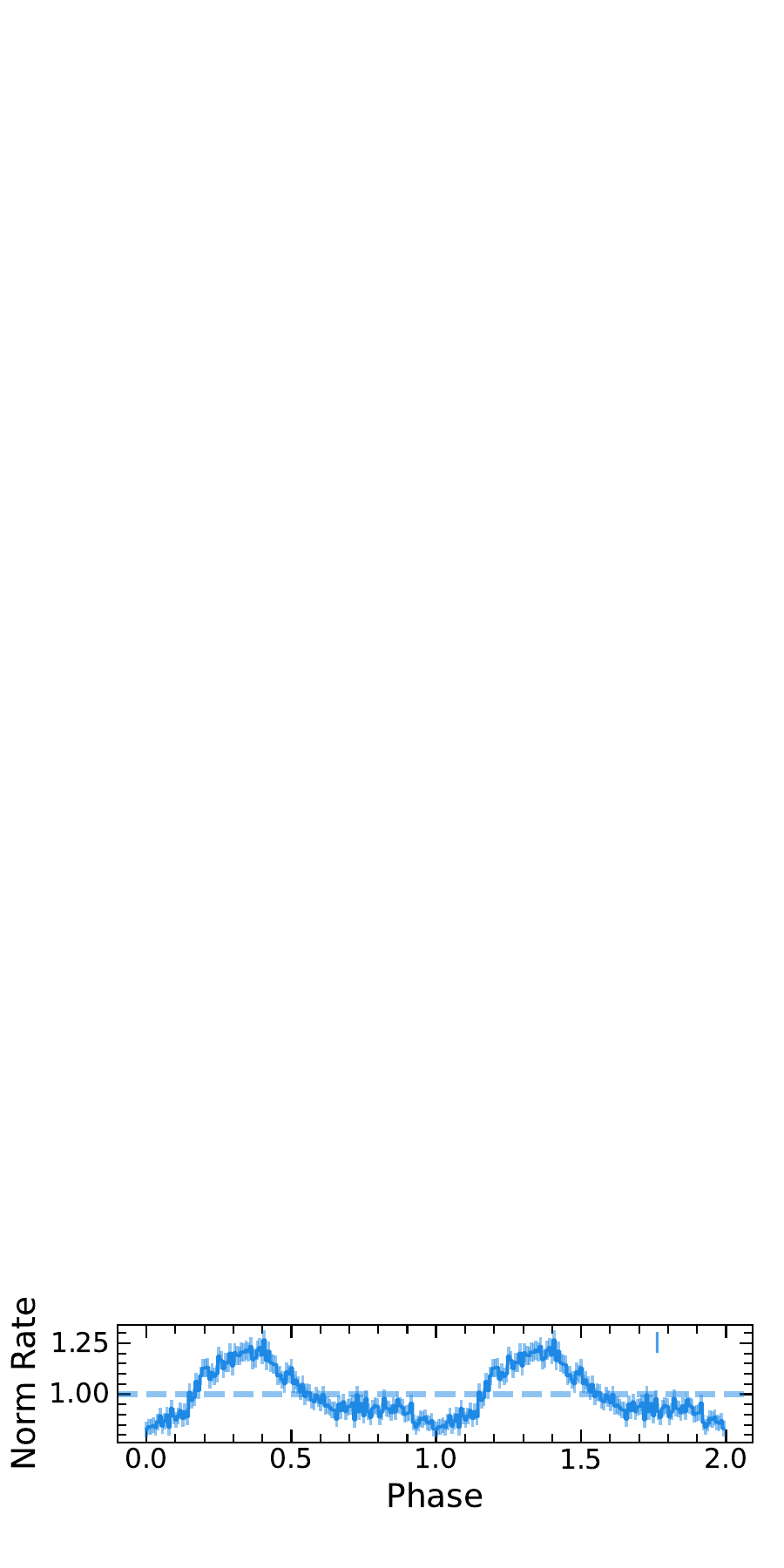}
    \includegraphics[width=.33\textwidth]{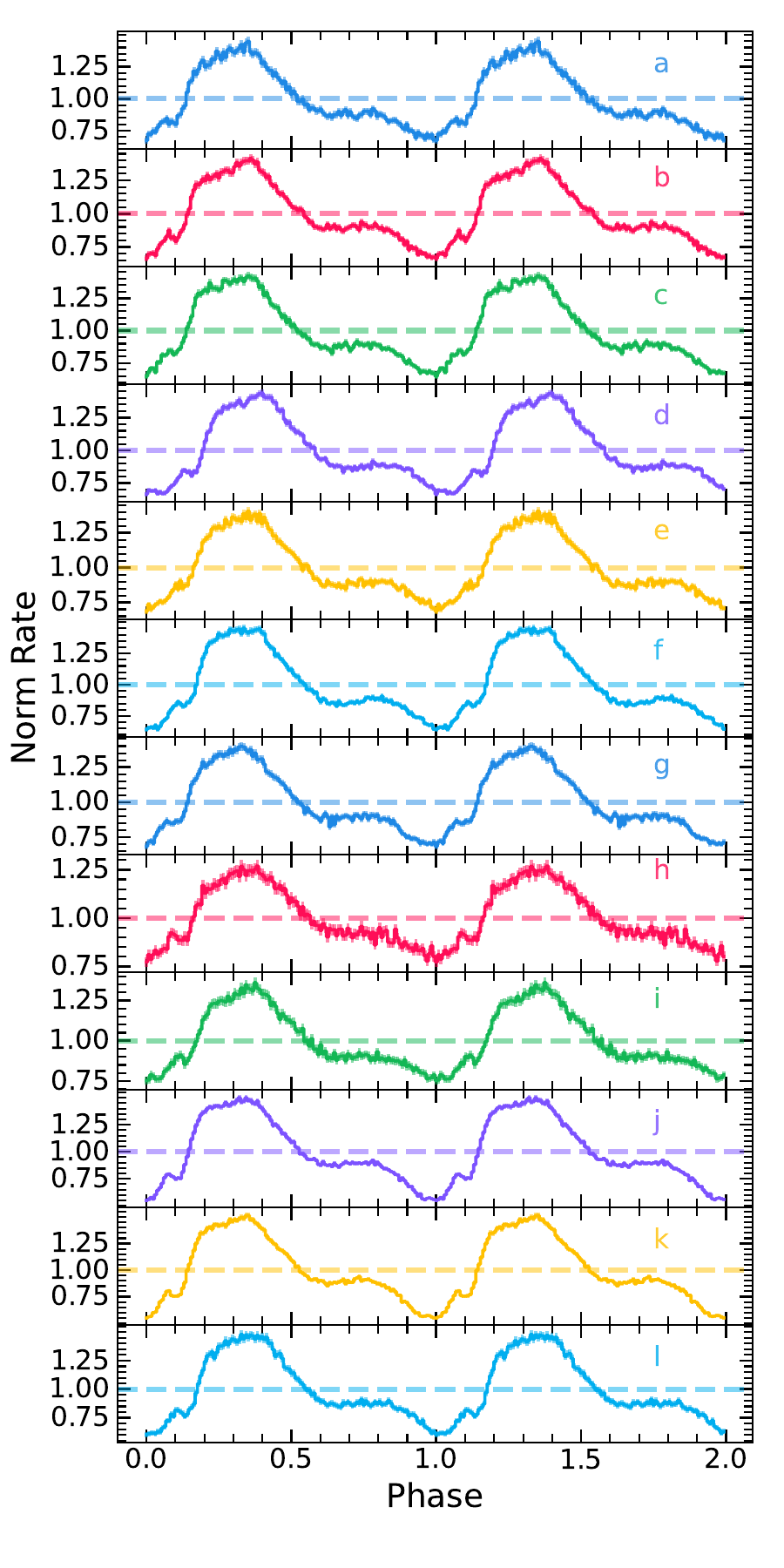}
    \includegraphics[width=.33\textwidth]{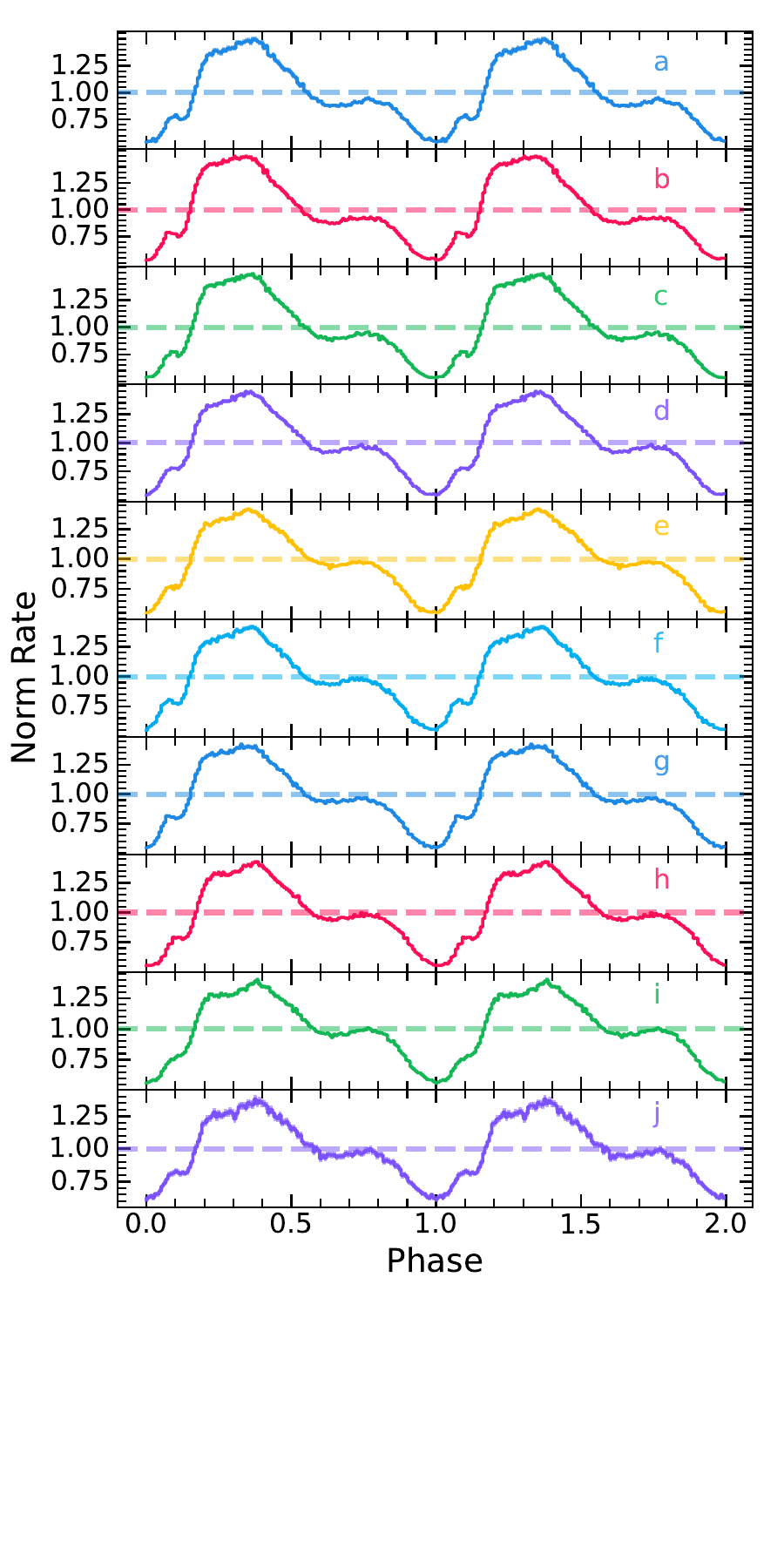}
    \caption{Pulse profiles of Cen X-3 in 3-78 keV from the summations of the FPMA and FPMB counts. These profiles are aligned by defining the minimum as phase zero. The first panel is from the first time interval. The second panel comes from  intervals 2 to 13 in the first orbit with a moderate luminosity. The third panel comes from intervals 14 to 23 in the second orbit with high luminosity.}
    \label{fig:pulse}
\end{figure*}

\begin{figure}
    \centering
    \includegraphics[width=.5\textwidth]{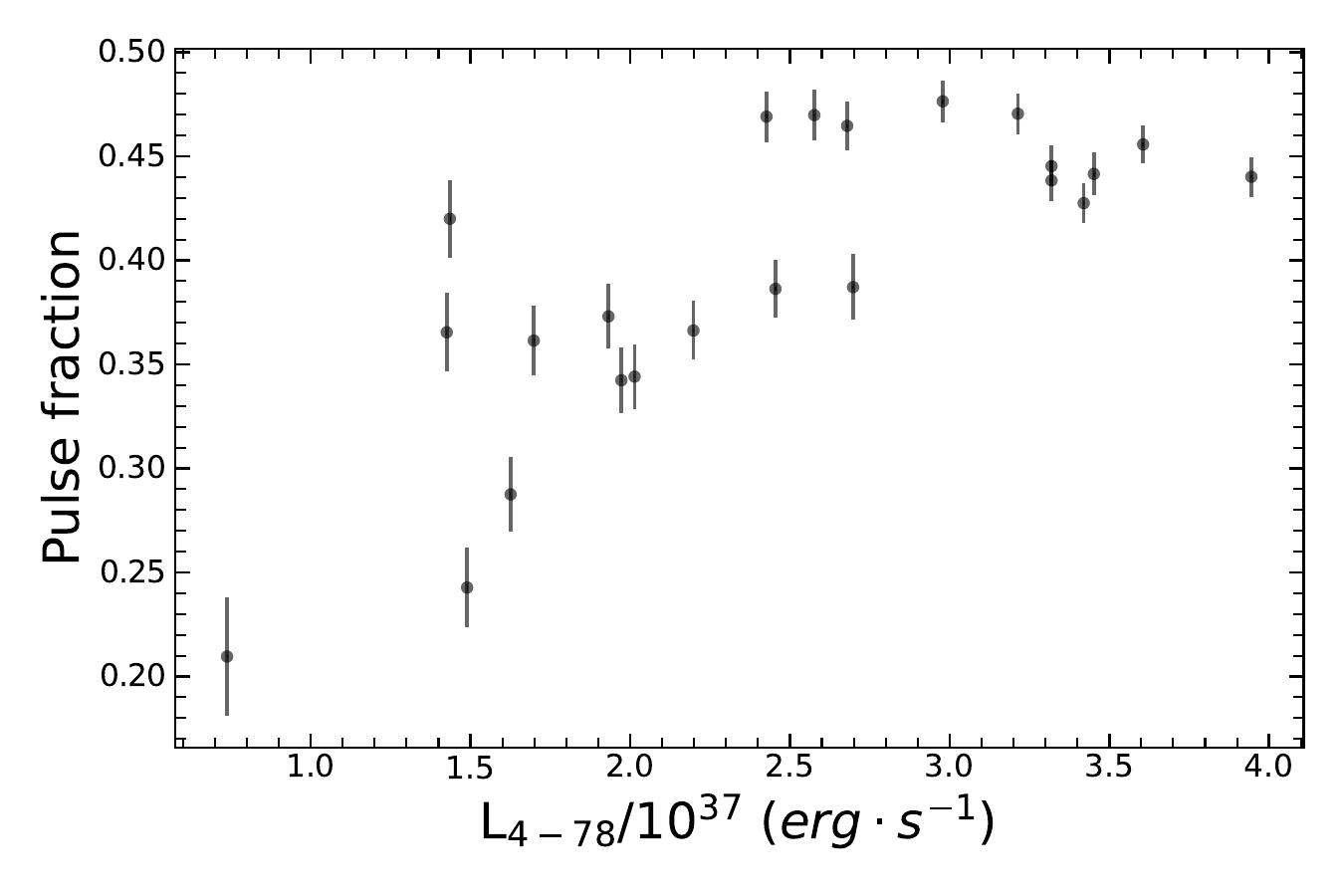}
    \caption{Pulse fraction (PF=($I_{max}$-$I_{min}$)/($I_{max}$+$I_{min}$)) in energy bands of 3-78 keV of Cen X-3 as a function of luminosity.}
    \label{fig:pf}
\end{figure}

\begin{figure}
    \centering
    \includegraphics[width=.5\textwidth]{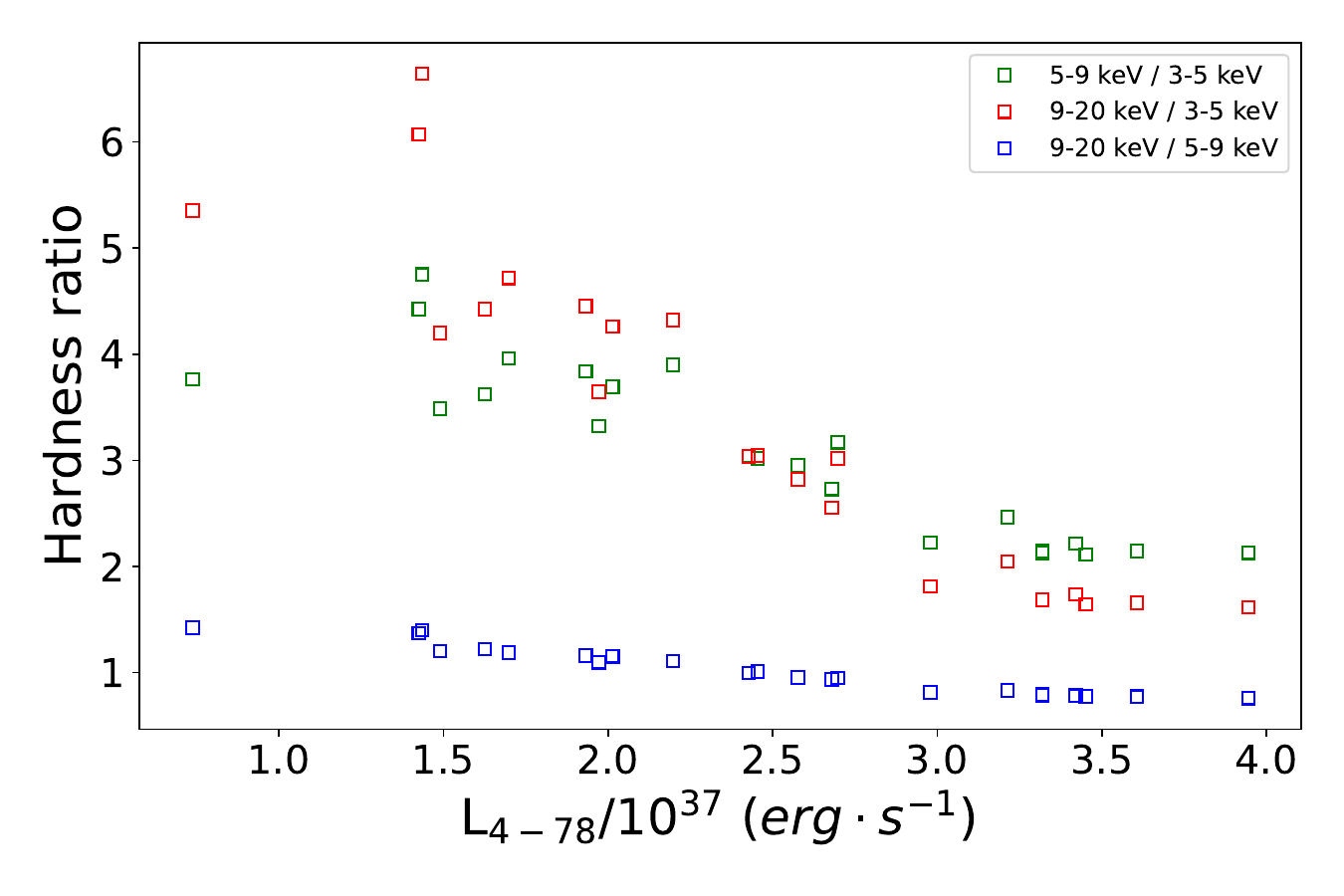}
    \caption{Hardness ratios of Cen X-3 as a function of the 4-78 keV luminosity in different energy bands.}
    \label{fig:hardness}
\end{figure}

To find the spin period of the Cen X-3 at each time interval, we searched for the pulsations using the epoch folding technique with the \textit{efsearch} tools. The best pulse period can be determined using the maximum $\chi^{2}$ value. The pulse profile was generated by folding each light curve by the best spin period. We plotted the pulse profiles in 3-78 keV energy bands for 23 time intervals. As seen in Figure \cref{fig:pulse}, the first panel is the pulse profile from the first time interval (here the middle time at MJD 59591.2); the second panel comes from intervals 2 to 13 in the first orbital phases ($\sim$ MJD 59591.7-59593.3) with a moderate luminosity; and the third panel comes from intervals 14 to 23 over the second orbit ($\sim$ MJD 59593.8-59595.1) with high luminosity. We find that the shape of pulse profile did not evolve over the orbital phase and remained stable. These pulse profiles show obvious double-peaked shapes, sometimes with a narrow and small peak at the pulse phase of $\sim$0.1 in the rising phases of the main pulse. There is no significant evolution of the pulse profile within one orbit, but the second or faint peak of the pulse in the second orbit appears to be stronger than the first orbit, maybe due to the higher luminosity. We also present the pulse fraction versus the corresponding flux and find that the pulse fraction increases as the flux rises, as seen in Figure \cref{fig:pf}. Figure \cref{fig:hardness} shows the hardness ratio in different energy bands over the 4-78 keV luminosity. The hardness ratios of different energy ranges show a negative correlation with the total luminosity, and at a higher rate (luminosities >$3 \times 10^{37}$ erg s$^{-1}$) they change, decreasing slowly before reaching a plateau where they remain stable.

\begin{figure*}
    \centering
   \includegraphics[width=.24\textwidth]{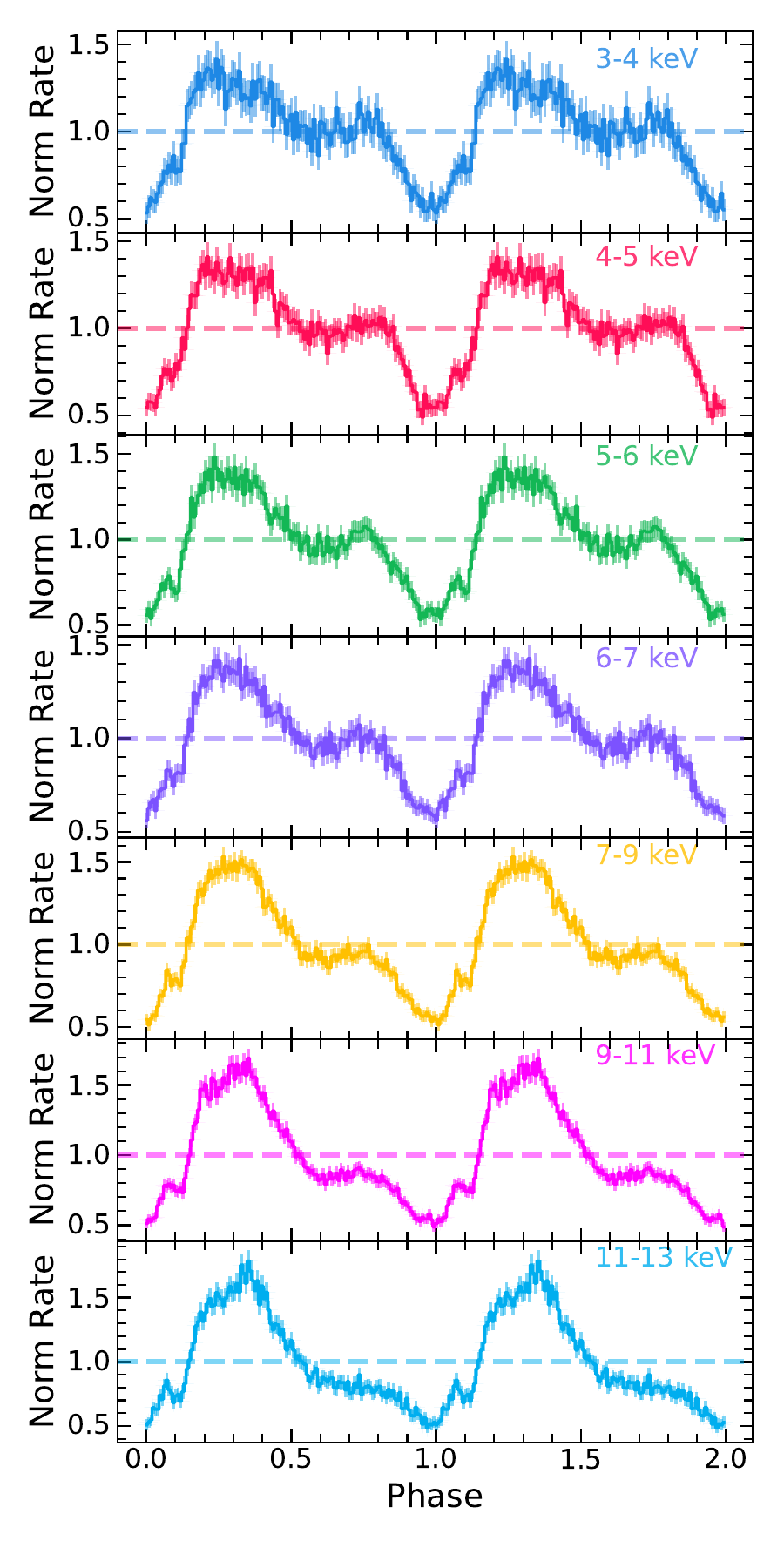}
   \includegraphics[width=.24\textwidth]{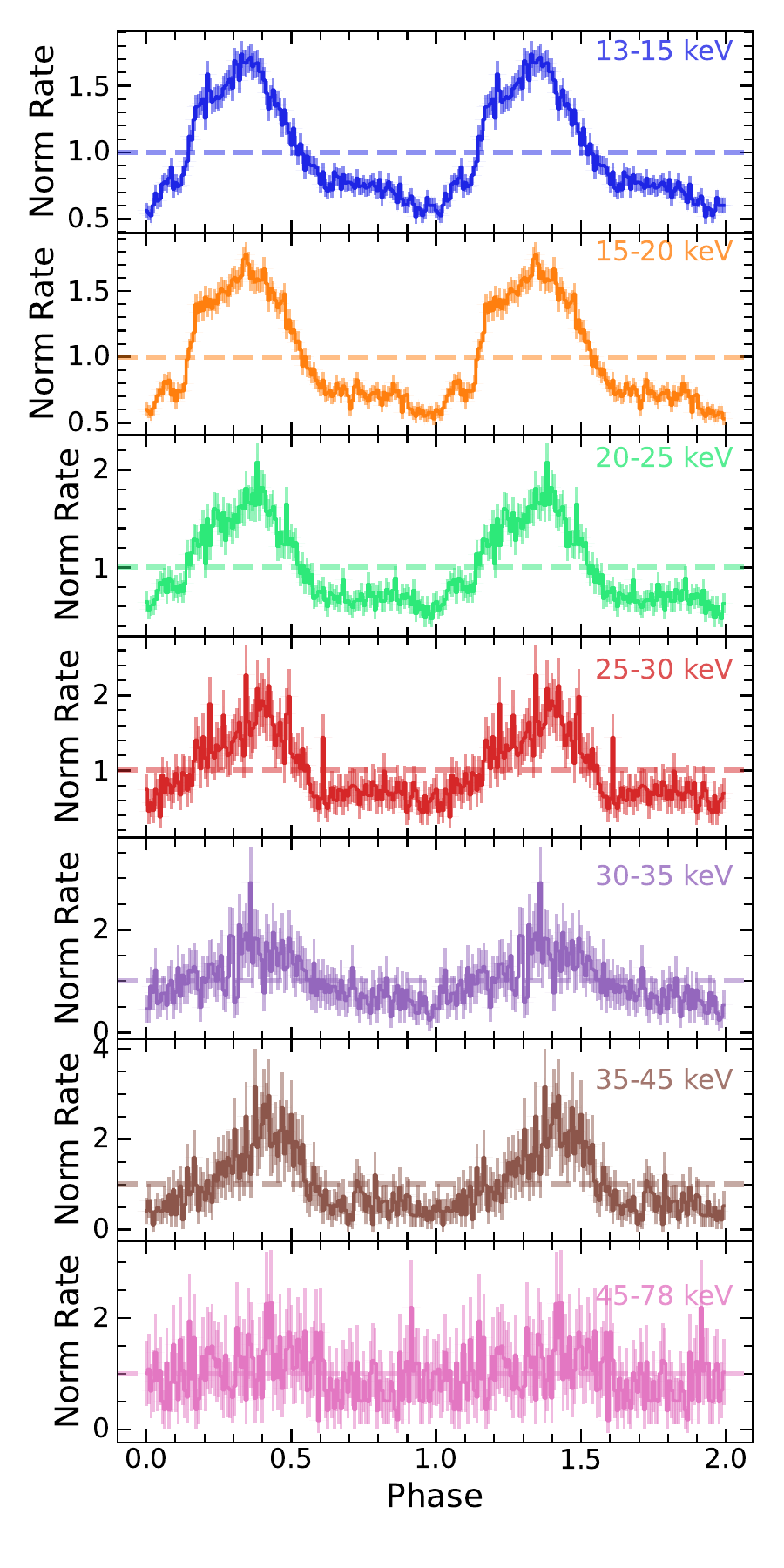}
   \includegraphics[width=.24\textwidth]{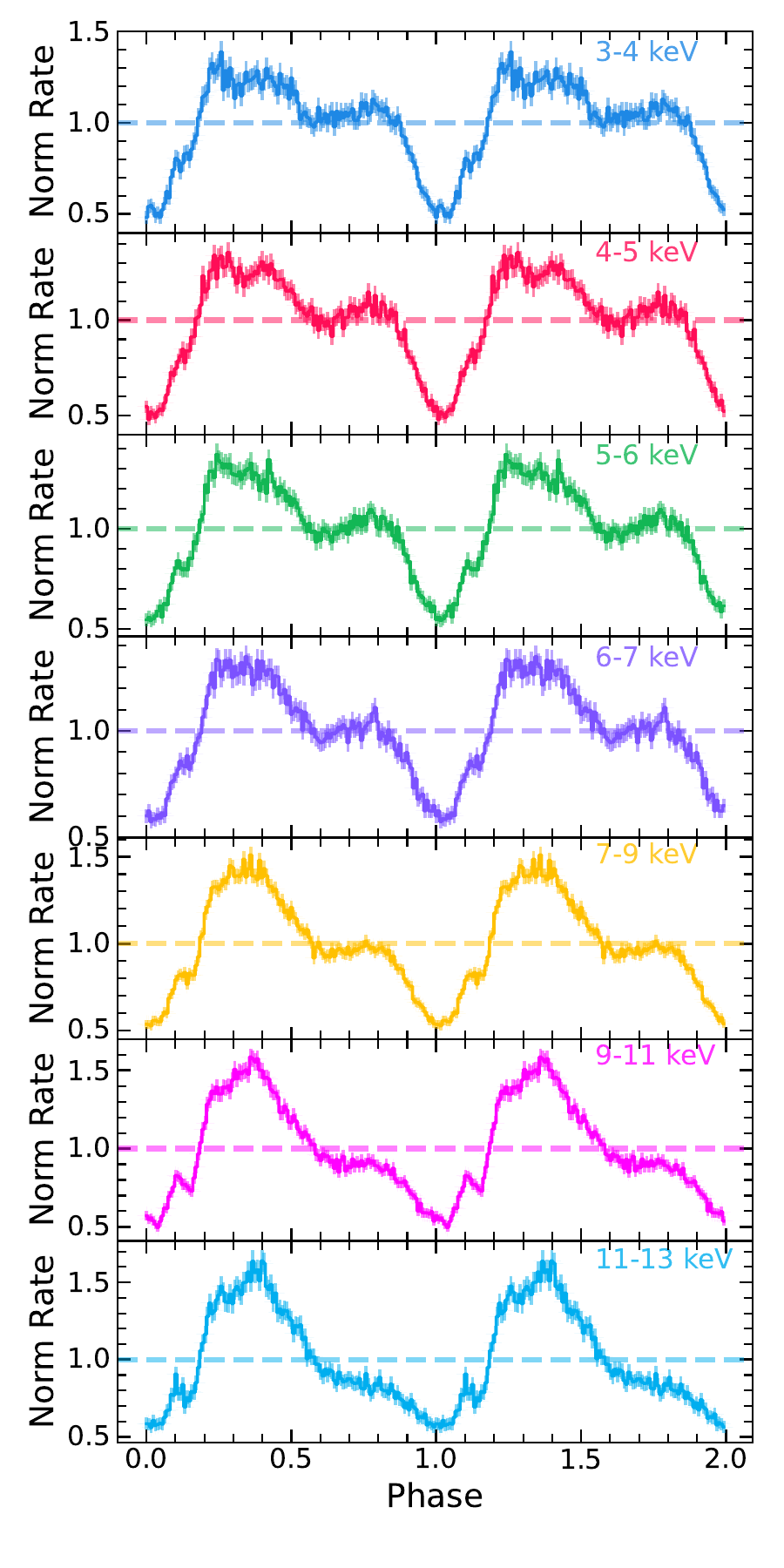}
   \includegraphics[width=.24\textwidth]{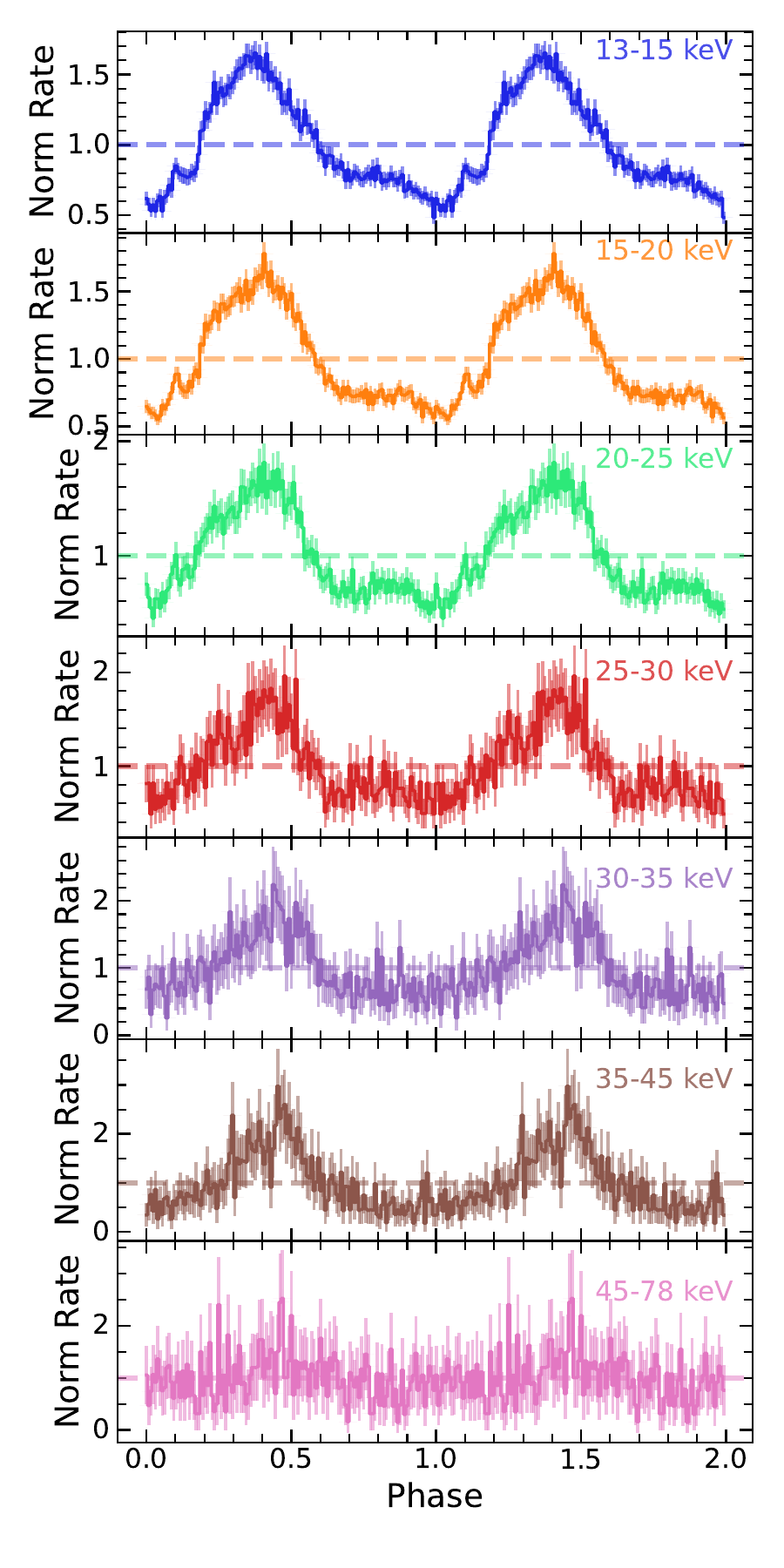}
    \caption{The pulse proﬁles of Cen X-3 in diﬀerent energy bands. The left two panels show the energy-resolved pulse profiles for Cen X-3 from the summations of the FPMA and FPMB counts in MJD 59593.12 (the time interval k of the first orbit) at low intensity state and the right two panels show the energy-resolved pulse profiles in MJD 59594.67 (the time interval g of the second orbit) at high intensity state.}
    \label{fig:pulse_profile}
\end{figure*}

To further investigate the energy-resolved pulse profiles, we focused on the time interval k of the first orbit (MJD 59593.12) at low states and another interval g of the second orbit (MJD 59594.67) at high states, and then folded the light curves into 14 different energy bands of 3-4 keV, 4-5 keV, 5-6 keV, 6-7 keV, 7-9 keV, 9-11 keV, 11-13 keV, 13-15 keV, and 15-20 keV, 20-25 keV, 25-30 keV, 30-35 keV, 35-45 keV, and 45-78 keV. Figure \cref{fig:pulse_profile} shows the energy-dependent pulse profiles over different energy bands for the two time intervals: that is, k in the low state (left two panels) and g in the high state (right two panels). Only the energy bands below 45 keV show clear pulsations. The double-peaked profile was detected in a low energy range of 3-15 keV, and the secondary peak decreases as energy increases. At a higher energy range above 15 keV, the pulse has a single-peaked shape. An additional narrow peak in the pulse phase of $\sim$ 0.1 at the rising stages of the main pulse is seen below 20 keV. Similar energy-dependent pulse profiles have been reported \citep{1992ApJ...396..147N,2000ApJ...530..429B,2008ApJ...675.1487S,2021JApA...42...58S,2021MNRAS.500.3454T,2023JHEAp..38...32L}. 

\begin{figure}
    \centering
    \includegraphics[width=.5\textwidth]{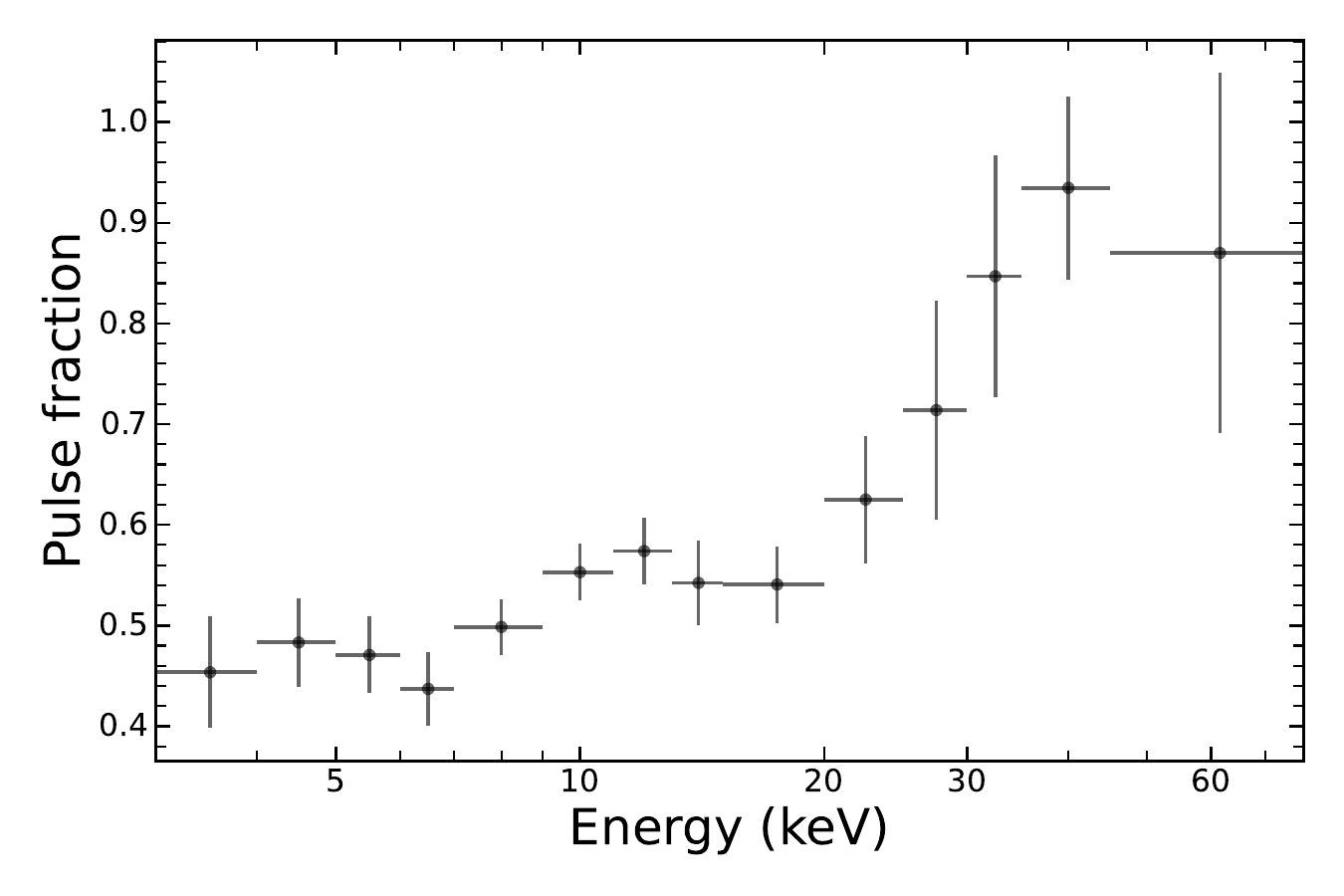}
    \includegraphics[width=.5\textwidth]{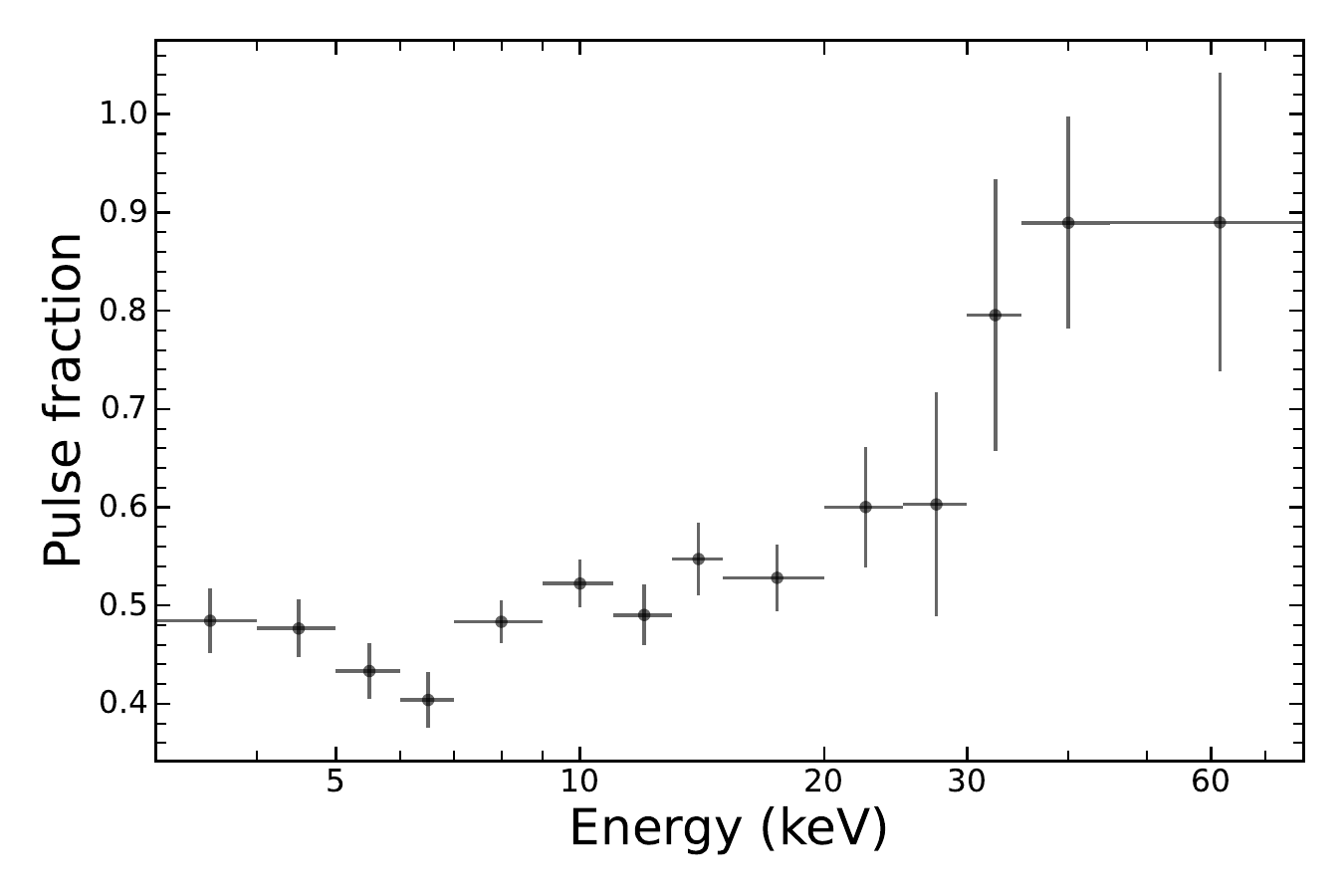}
    \caption{Pulse fraction as a function of energy in 59593.12 MJD and 59594.67 MJD, respectively.}
    \label{fig:pf_energy}
\end{figure}

In addition, we find multiple subtle peaks in the pulse profile in the energy band of 3-5 keV at a high intensity state. Also, the main peak is divided into two subpeaks, which is particularly different from the behavior of the pulse profile seen in previous observations. The evolution from a double-peaked profile at low states to a triple-peaked profile at high states may indicate the transition from a subcritical to a critical regime. It is also worth mentioning that the maximum of the main pulse at the high states in the energy band of 3-5 keV shows an obvious shift of about 0.1 compared to other energy ranges, which may indicate the emissions come from both "pencil" and "fan" beams. Regarding the asymmetric shape of the pulse profile, this may be explained as being due to contributions from two emission regions displaced from the dipole geometry by approximately 10$^\circ$, as proposed by \cite{1996ApJ...467..794K}. Recently, X-ray polarization research on Cen X-3 performed by the Imaging X-ray Polarimetry Explorer (IXPE) \citep{2022ApJ...941L..14T} showed that the pulsar geometry parameters and the phase-resolved polarimetric properties are in good agreement with the pulse-profile decomposition by \cite{1996ApJ...467..794K}. \cite{2024arXiv240101226S} introduced the phase-correlated variability analysis (PCVA) method and successfully decomposed the X-ray pulsar profiles of Cen X-3, but their results can produce asymmetric pulses, which are not aligned with those of \cite{1996ApJ...467..794K}, who assume that the emissions at the two poles are symmetrical. We expect the results of \cite{2024arXiv240101226S} {could explain the observed splitting} of the main peak in our study. The complex emission profile revealed by our analysis is probably a mixture of the pencil and fan beam pattern. \cite{Tamba_2023} suggested that a double-peaked shape at low energy and a single-peaked shape at high energy may indicate the presence of both pencil and fan beam emissions. Thus at higher energy states, the beam patterns probably become dominated by fan beam and a triple peak is expected, which is consistent with our results at low energy. Figure \cref{fig:pf_energy} shows the energy-dependent pulse fraction ---{defined here by the expression ($I_{max}$-$I_{min}$)/($I_{max}$+$I_{min}$)--- in energy bands from 3 to 78 keV} for the two intervals above. In the range of 3–78 keV, the pulse fraction in both states increases as the energy increases, except for the band of 6–7 keV, which corresponds to the 6.4 keV iron emission line.

\section{Spectral analysis}\label{sec:result}

The X-ray spectra in Cen X-3 are usually characterized by a power law modified by {an exponential cut-off}, the iron emission lines, a soft excess, and the CRSF around 30 keV. For the X-ray spectral fitting, there are some analytic functions that are commonly used to describe the broadband continuum. In order to obtain a suitable or good description of the data, we used these empirical models and compared the fits among them. First, we applied a common continuum spectrum of a power law with a Fermi-Dirac cutoff (FDCUT) ---also adopted by \cite{2008ApJ...675.1487S}, \cite{Tamba_2023}, and \cite{LIU2023}--- to fit the spectra in Cen X-3, which is characterized by
\begin{equation}
C(E)=N \cdot E^{-\Gamma}\left[1+\exp \left(\frac{E-E_{\text {cutoff }}}{E_{\text {fold }}}\right)\right]^{-1},
\end{equation}
where $\Gamma$ is the power-law index, and $E_{\text {cutoff }}$ and $E_{\text {fold }}$ are the cutoff and folding energies, respectively. {Another choice is the power law with a high-energy cutoff (HIGHECUT) previously employed for Cen X-3 \citep{1983ApJ...270..711W,2000ApJ...530..429B}, which has the form
\begin{equation}
C(E)=\left\{\begin{array}{ll}
N E^{-\Gamma} & \text { for } E \leq E_{\text {cutoff }}, \\
N E^{-\Gamma} \times \exp \left[-\left(E-E_{\text {cutoff }}\right) / E_{\text {fold }}\right] & \text { for } E>E_{\text {cut }} .
\end{array}\right.
\end{equation}
Although C(E) has a discontinuous derivative around $E_{\text {cutoff }}$, which can result in wedge-shaped residuals around 15 keV, it is extensively used and adequate for modeling spectra. Finally, the sum of positive and negative power laws multiplied by an exponential cutoff (NPEX) \citep{Mihara1995} is expressed as
\begin{equation}
C(E)=\left(N_1 \cdot E^{-\Gamma_1}+N_2 \cdot E^{+\Gamma_2}\right) \exp \left(-E / E_{\mathrm{fold}}\right),
\end{equation}
where $\Gamma_2$ is usually set to 2.} The photoelectric absorption $f(N_{\rm H}^{\rm IM})$ (phabs in XSPEC) component and iron line emission feature around 6.4 keV $f_{\rm gauss}$ (Gaussian in XSPEC) are also included in our models, but the soft excess is not included in the model because this contribution can be neglected. As the hydrogen column density $N_{\rm H}^{\rm IM}$ for the interstellar medium cannot be constrained well in the energy above 4 keV, we fixed it to 1.1 $\times 10^{22}$ atoms cm$^{-2}$ according to the galactic hydrogen column density in the same direction as the source. For the additional absorption $f(N_{\rm H})$ for the stellar wind or the local materials, we tried the photoelectric absorption (phabs) and partial covering absorption (pcfabs in XSPEC) models {using the abundance of {\tt angr} \citep{1989GeCoA..53..197A} and {\tt vern} cross-sections \citep{1996ApJ...465..487V}. The $\chi^2$ fitting statistics are almost the same for the two absorption models applied to the NPEX continuum. However, for the HIGHECUT and the FDCUT continua, the pcfabs model performs better, especially at low luminosities (< $\sim 3 \times 10^{37}$ erg s$^{-1}$). At higher luminosities, the pcfabs parameters are not well constrained for the FDCUT model in some cases, and so we used the phabs absorption models instead.} 

To model the CRSF, we used a multiplicative absorption line model with a Gaussian optical depth profile (gabs in XSPEC):
\begin{equation}
gabs(E)=\exp \left(-\frac{d_{\mathrm{cyc}}}{\sqrt{2 \pi} \sigma_{\mathrm{cyc}}} e^{-0.5 \left[\left(E-E_{\mathrm{cyc}}\right) / \sigma_{\mathrm{cyc}}\right]^2}\right) ,
\end{equation}
where $E_{\mathrm{cyc}}$ is the cyclotron line energy,   $d_{\mathrm{cyc}}$ is the line depth, and $\sigma_{\mathrm{cyc}}$ is the width.  We also leave the normalization constant free to vary between the FPMA and FPMB spectra. Therefore, the total model adopted here is
\begin{equation}
    S(E)= constant \times f({\rm N_H^{IM}}) \times f({\rm N_H }) \times (C(E) * gabs(E) + f_{\rm gauss})
.\end{equation}
The spectra were fitted in the energy range of 4–78 keV. We grouped the spectra from both FPMA and FPMB with a minimum of 25 counts per bin.

\begin{figure}
    \centering
    \includegraphics[width=.49\textwidth]{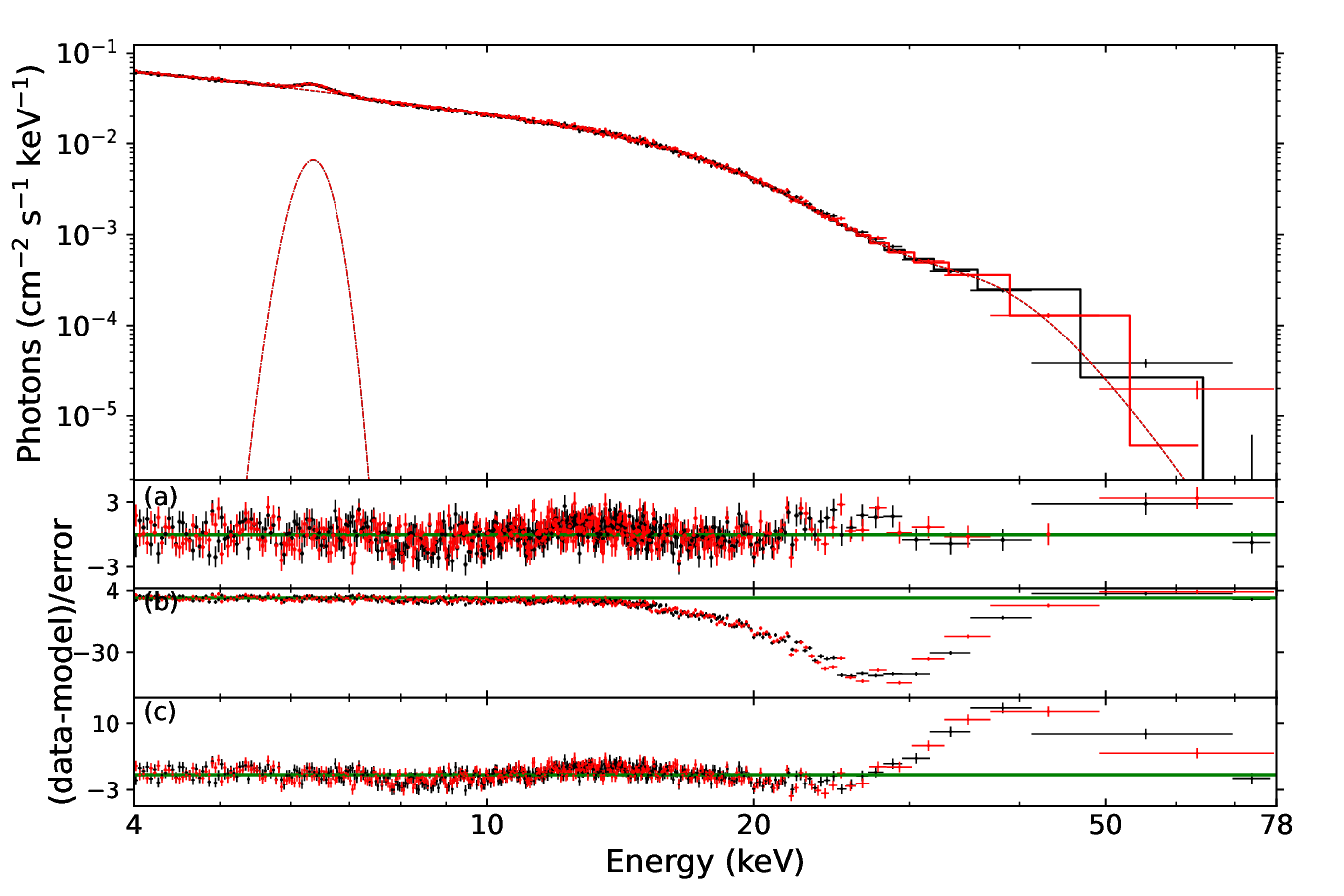}
    \caption{ X-ray spectrum of Cen X-3 ranging from 4–78 keV for the interval g of the second orbit using the Fermi-Dirac cutoff continuum. {Panel (a): Residuals of the best-fitting model. Panel (b): Residuals of the best model, setting the strength of the 30 keV cyclotron line to zero. Panel (c): Residuals of the spectral fitting without the CRSF.}}
    \label{fig:spectrum}
\end{figure}

\begin{table}[!htp]
\centering
\caption{All the best-fitting parameters for the specific time interval g of the second orbit (59594.67 MJD) of Cen X-3 observed with NuSTAR in 2022. The flux is given in units of erg cm$^{-2}$ s$^{-1}$.}
\renewcommand\arraystretch{1.25}
\setlength{\tabcolsep}{2.1mm}{
\begin{tabular}{lrrr}
\hline
Parameter & value$^1$ & value$^2$ & value$^3$  \\

\hline
 $E_{\rm cyc}$ (keV) & $27.91_{-0.44}^{+0.37}$  & $28.92_{-0.57}^{+0.83}$ & $28.35_{-0.44}^{+0.57}$  \\
 $W_{\rm cyc}$ (keV) & $6.48_{-0.26}^{+0.21}$ & $4.27_{-0.39}^{+0.72}$  & $6.92_{-0.41}^{+0.61}$ \\
 $D_{\rm cyc}$ & $19.96_{-2.43}^{+2.25}$ & $4.48_{-0.62}^{+1.37}$ & $15.43_{-1.98}^{+3.29}$  \\
 $N_{\rm H}^{\rm IM}$  & $1.1$  & $1.1$ &  $1.1$ \\
 $N_{\rm H}$  & $1.88_{-0.41}^{+0.44}$ & $3.92_{-0.41}^{+1.14}$  & $2.98_{-0.59}^{+0.67}$  \\
 CvrFract  & -  & $0.95_{-0.20}^{+0.02}$  & -  \\
 $\Gamma$ & $1.31_{-0.02}^{+0.02}$  & $1.48_{-0.03}^{+0.02}$  & $0.85_{-0.04}^{+0.06}$  \\
 norm &  $0.476_{-0.021}^{+0.027}$ &  $0.668_{-0.044}^{+0.042}$ &  $0.597_{-0.037}^{+0.048}$ \\
$E_{\rm cutoff}$(keV) & $26.69_{-1.06}^{+1.00}$ & $13.84_{-0.12}^{+0.10}$ & -  \\
$E_{\rm fold}$(keV) &  $4.96_{-0.28}^{+0.27}$ & $9.93_{-0.18}^{+0.28}$ & $4.39_{-0.06}^{+0.08}$  \\
 norm2 &  - & - & $0.0014_{-0.0001}^{+0.0001}$  \\
$E_{\rm Fe}$(keV) &  $6.36_{-0.03}^{+0.03}$  &  $6.38_{-0.02}^{+0.02}$ & $6.37_{-0.02}^{+0.02}$  \\
$\sigma_{\rm Fe}$(keV) & $0.25_{-0.04}^{+0.04}$  & $0.20_{-0.04}^{+0.05}$  & $0.27_{-0.05}^{+0.07}$    \\
$I_{\rm Fe}$(keV) & $0.0043_{-0.0005}^{+0.0008}$  & $0.0037_{-0.0004}^{+0.0007}$ & $0.0046_{-0.0006}^{+0.0009}$  \\
constant & $1.008_{-0.003}^{+0.004}$ & $1.008_{-0.004}^{+0.003}$ & $1.008_{-0.003}^{+0.003}$  \\
reduced $\chi^2$ (dof) & 1.097 (1324)  & 1.082 (1323)  & 1.038 (1324)  \\
$\rm Log \ Flux_{4-78}$ & $-8.233_{-0.001}^{+0.001}$ & $-8.219_{-0.001}^{+0.001}$ & $-8.225_{-0.001}^{+0.001}$   \\
\hline
\end{tabular}
}
\begin{tablenotes}
\item $^1$ Model: constant*phabs*pcfabs (or phabs)*(powerlaw  *fdcut*gabs +  gaussian).
\item $^2$ Model: constant*phabs*pcfabs*(powerlaw*highecut*gabs +  gaussian).
\item $^3$ Model: constant*phabs*phabs*(powerlaw*npex*gabs + gaussian).
\end{tablenotes}
\label{tab:pars}
\end{table}

As an example of a specific time {interval g of the second orbit} (59594.67 MJD), Figure \cref{fig:spectrum} shows the best-fitting spectra and the residuals between the spectra and the model {using the FDCUT continuum}. The residuals around $\sim$28 keV after removing the CRSF from the best fit are clearly seen in panel (b) of Figure \cref{fig:spectrum}. Panel (c) also shows the residuals of the spectral fitting without the CRSF, which cannot give good fits for the spectra, suggesting that the CRSF is also statistically required. The reduced $\chi^{2}$/d.o.f of the fitting without and with the gabs component are 1.88/1327 and 1.09/1324, respectively. Table \cref{tab:pars} lists all the best-fit parameters of this spectrum. The centroid energies of CRSFs for the best fitting are found at $27.91_{-0.44}^{+0.37}$ keV, and the iron emission line is around 6.4 keV. {Table \cref{tab:pars} also lists the fitting results for the HIGHECUT continuum and NPEX continuum models. We can see that the NPEX model is of slightly better quality (with a smaller reduced $\chi^{2}$) than the FDCUT and HIGHECUT models. The cyclotron line energy does not significantly change for these continuum models, except that the width and depth of the HIGHECUT model are smaller than those of the other two models.} 

\begin{figure*}
    \centering
    \includegraphics[width=.49\textwidth]{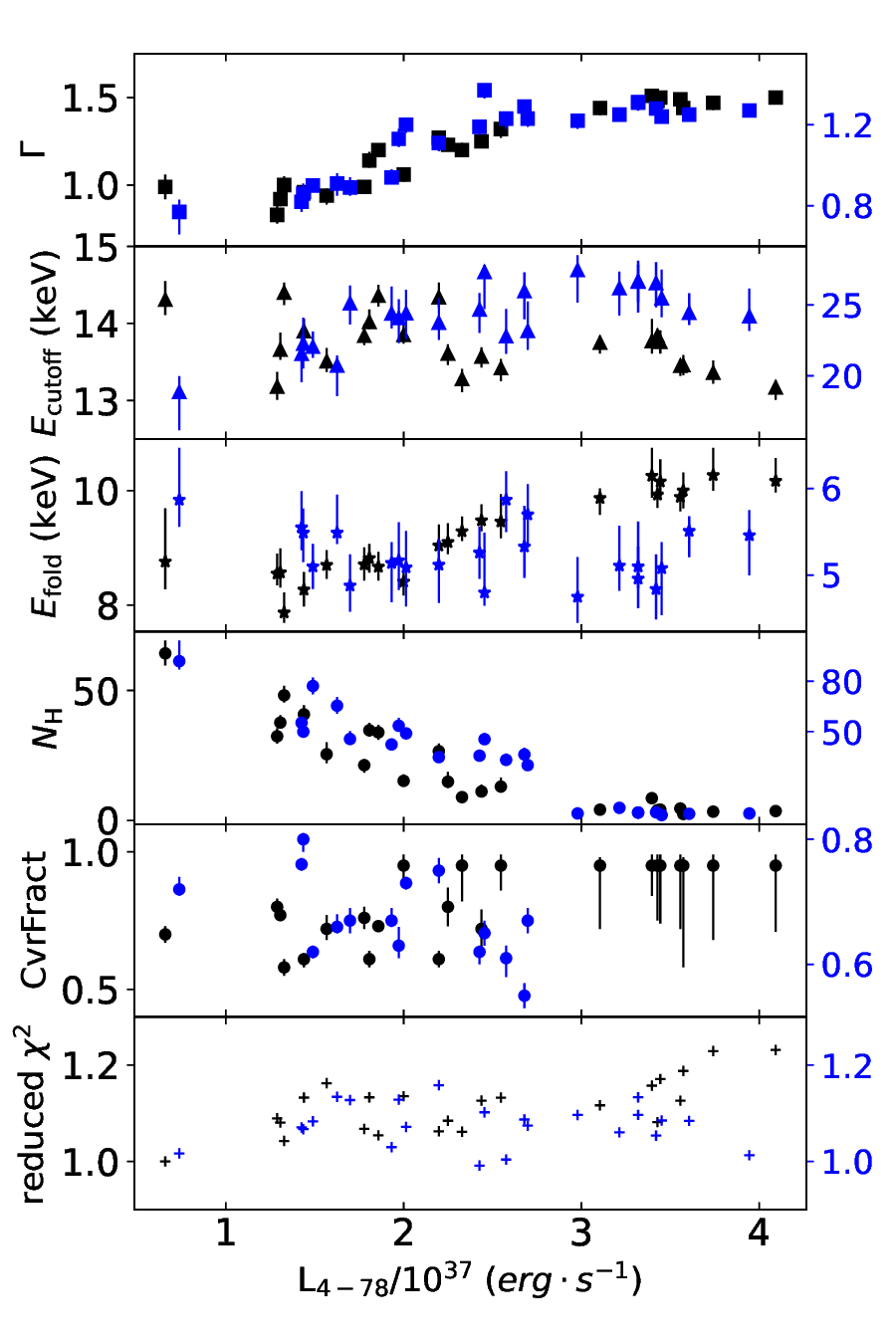}
    \includegraphics[width=.49\textwidth]{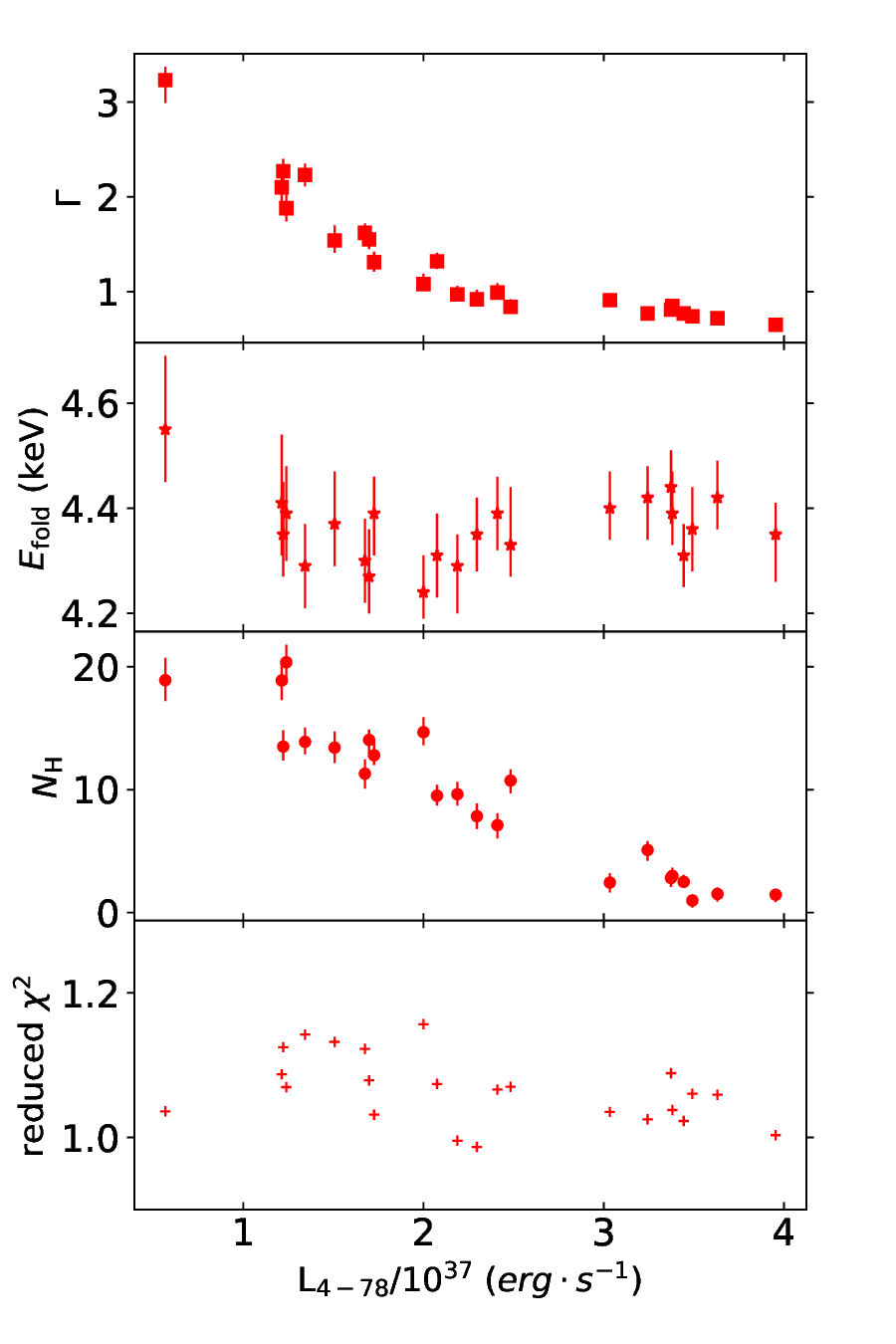}
    \caption{Spectral continuum parameters (photon index $\Gamma$, cutoff energy $E_{\rm cutoff}$, fold energy $E_{\rm fold}$, hydrogen column density $N_H$, and reduced $\chi^2$) as a function of X-ray {unabsorbed} luminosity. {The left plots are the fitting results from the FDCUT model (the blue points) and the HIGHECUT model (the black points), and the right plots show the parameters from the NPEX model.}}
    \label{fig:cont_Flux}
\end{figure*}

\begin{figure*}
    \centering
    \includegraphics[width=.49\textwidth]{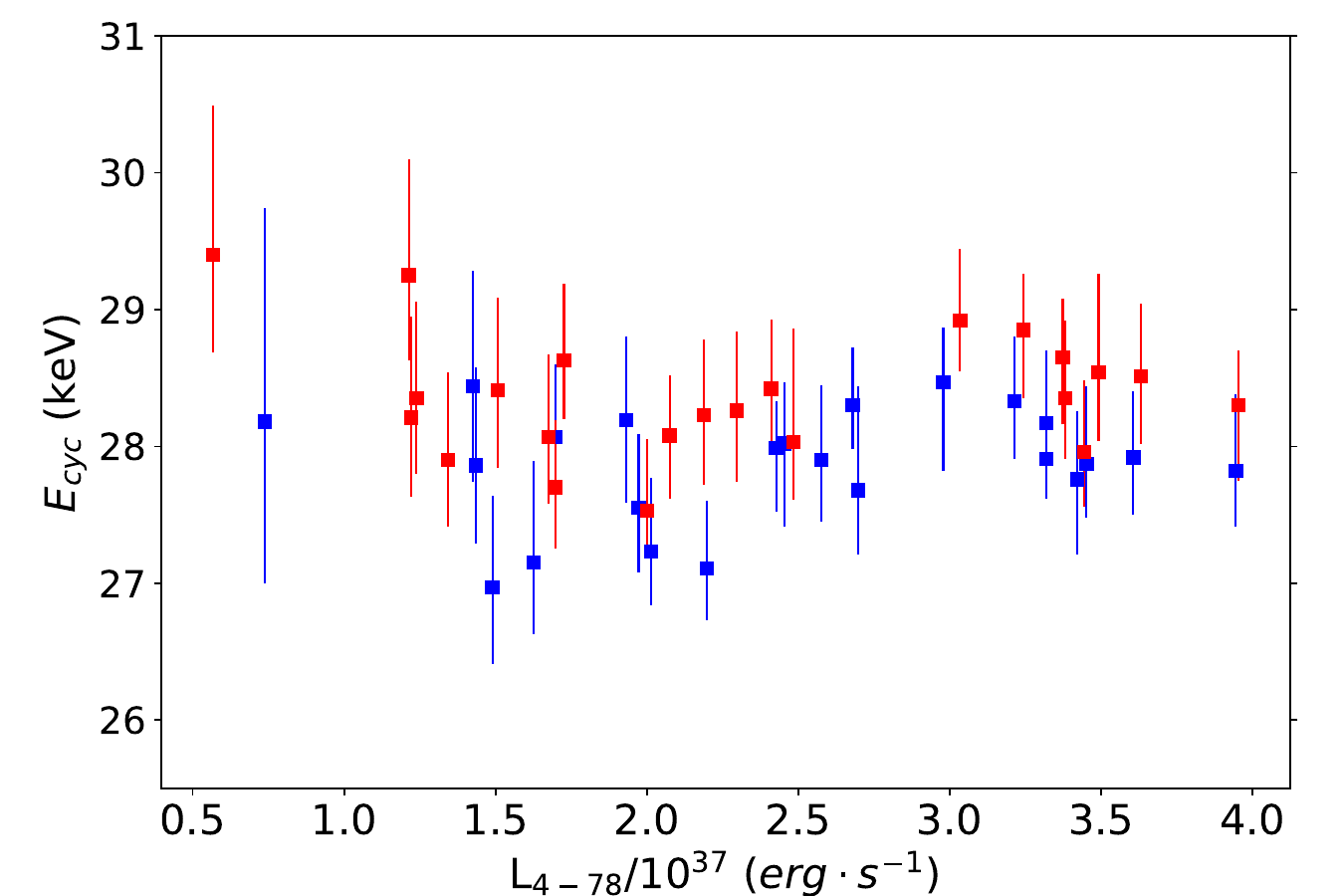}
    \includegraphics[width=.49\textwidth]{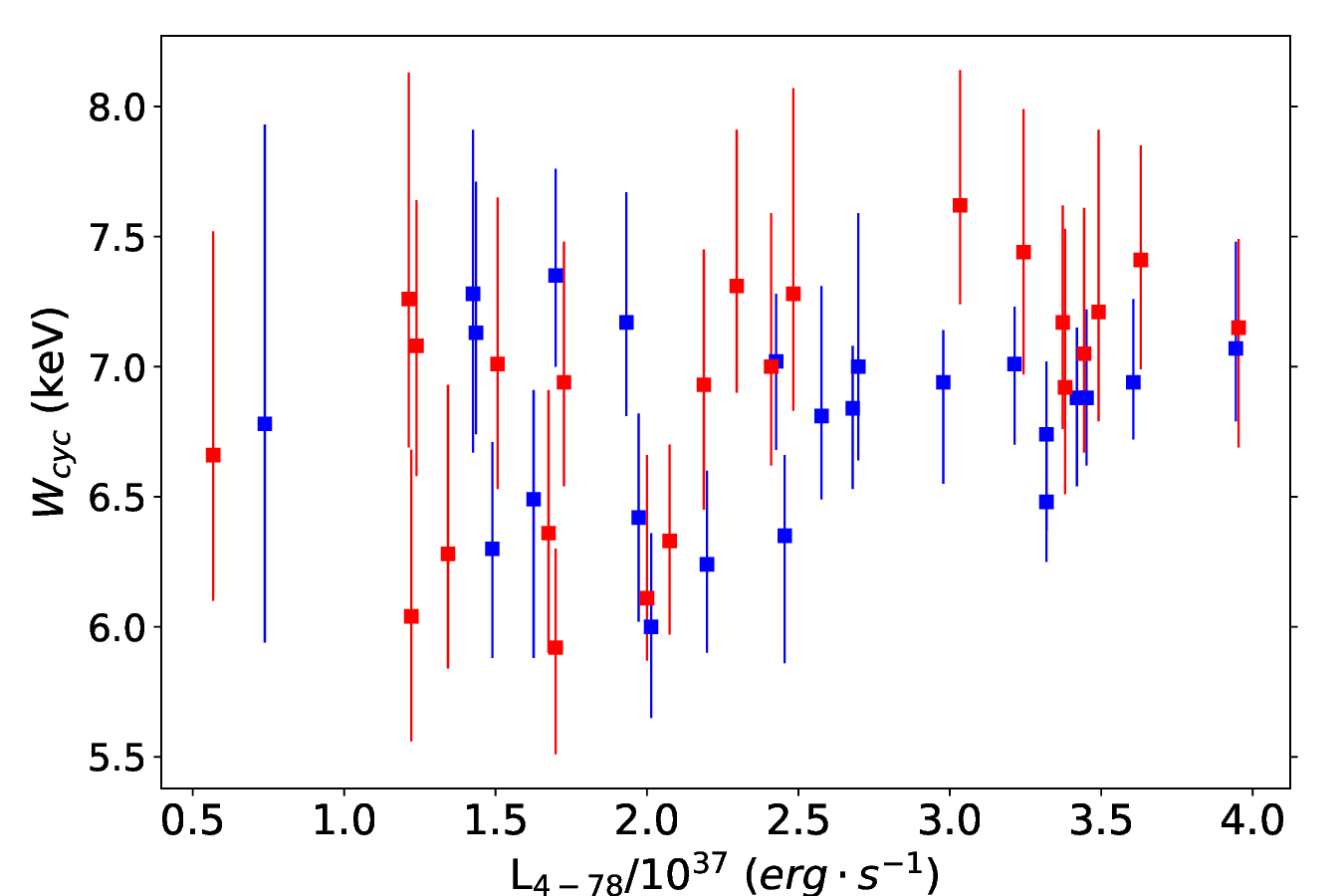}
    \includegraphics[width=.49\textwidth]{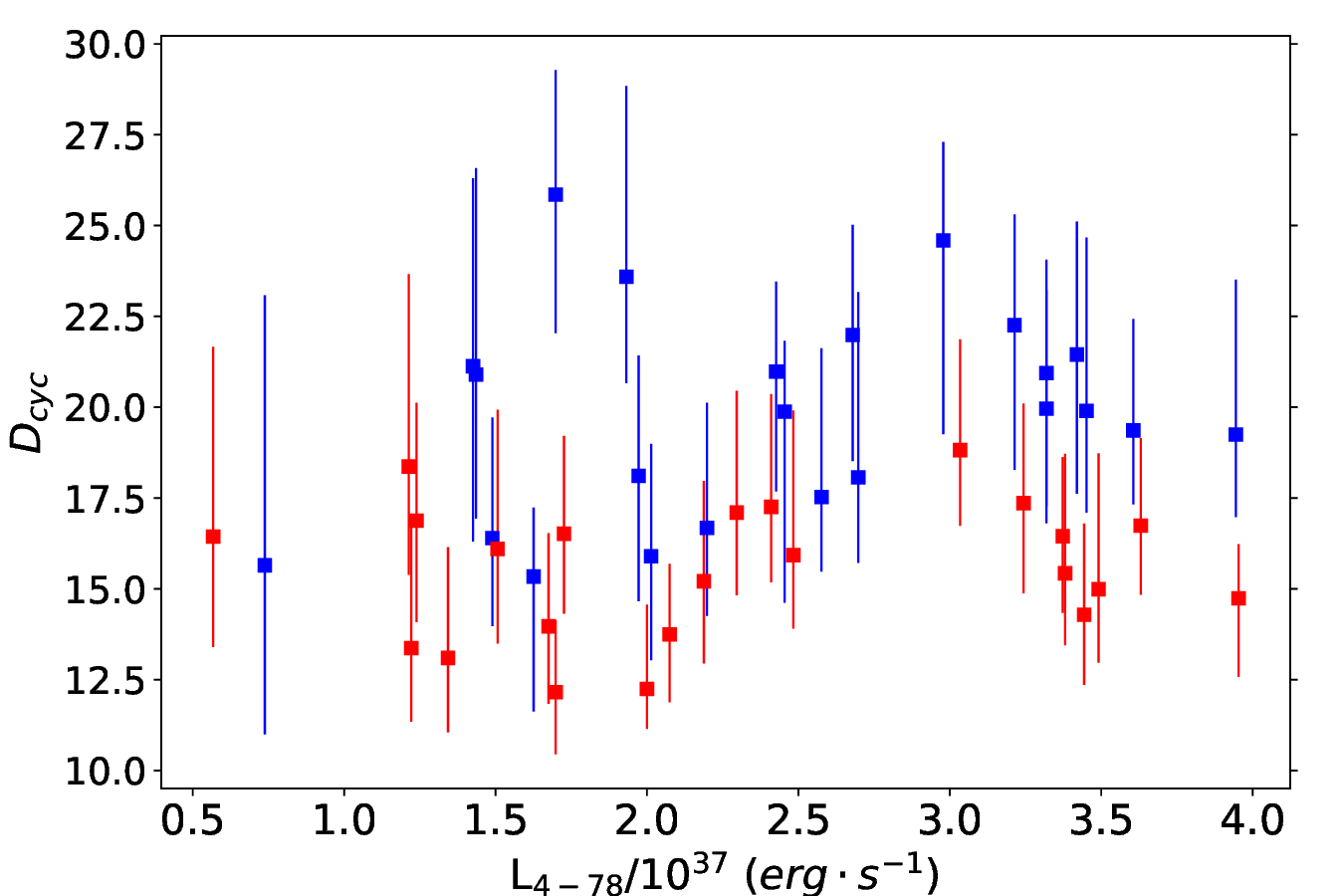}
    \includegraphics[width=.49\textwidth]{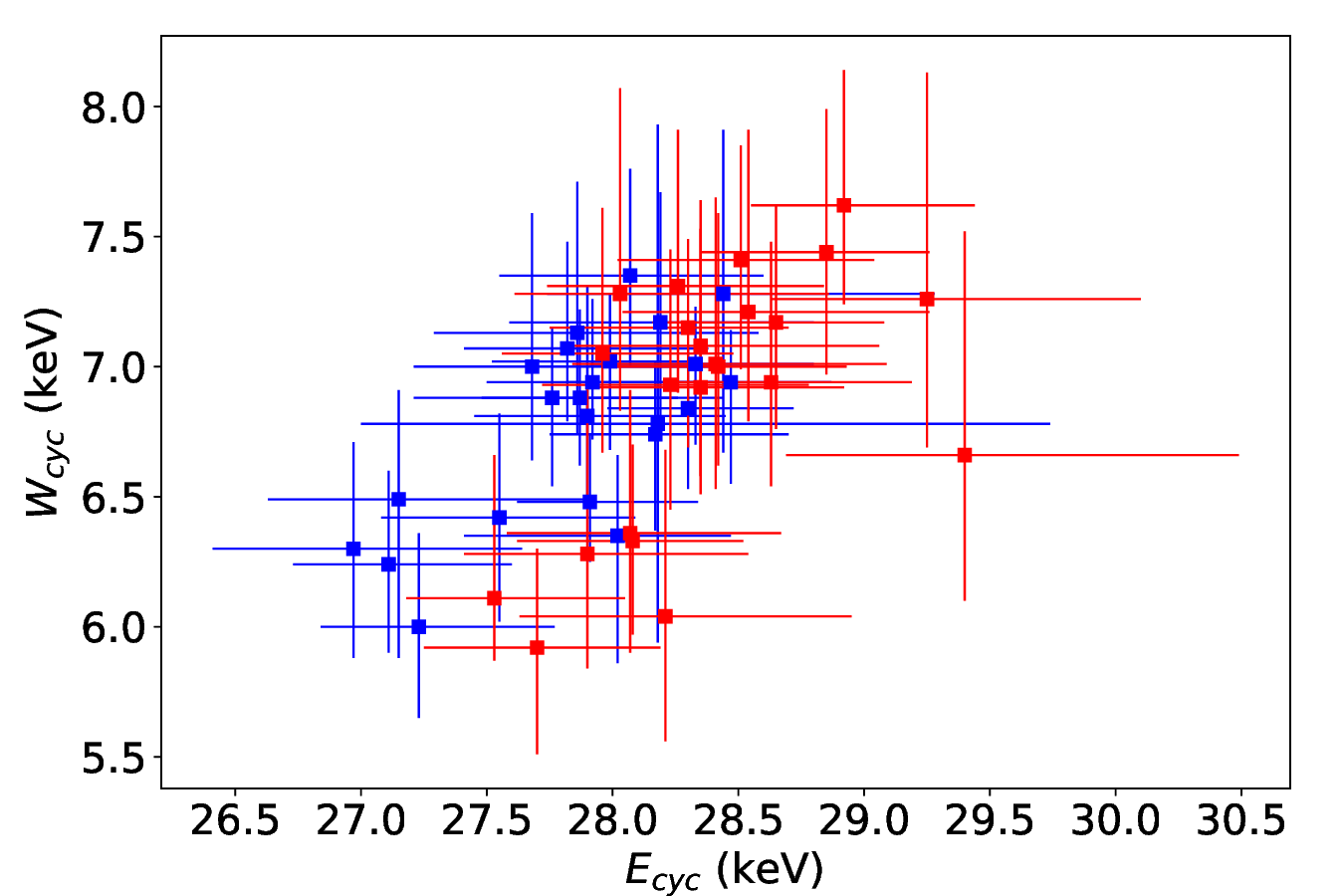} 
    \caption{ CRSF parameters (energy, width, and depth) as a function of X-ray luminosity and the centroid energy versus the width of CRSF. {These parameters are derived from the FDCUT model (the blue points) and the NPEX model (the red points), respectively.}}
    \label{fig:cyc_Flux}
\end{figure*}

\section{Results and discussion} \label{sec:discussion}

\begin{figure*}
    \centering
    \includegraphics[width=.49\textwidth]{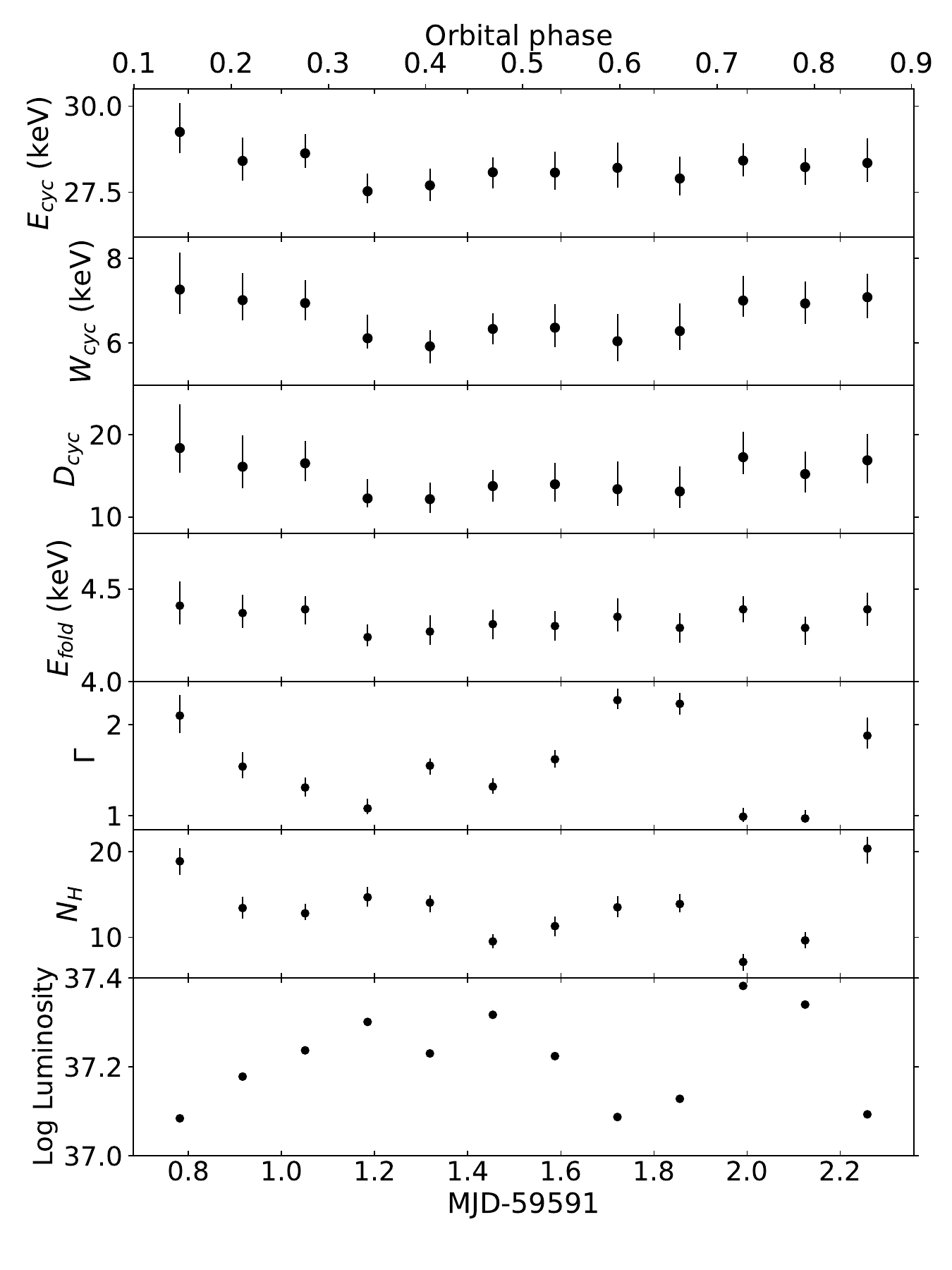}
    \includegraphics[width=.49\textwidth]{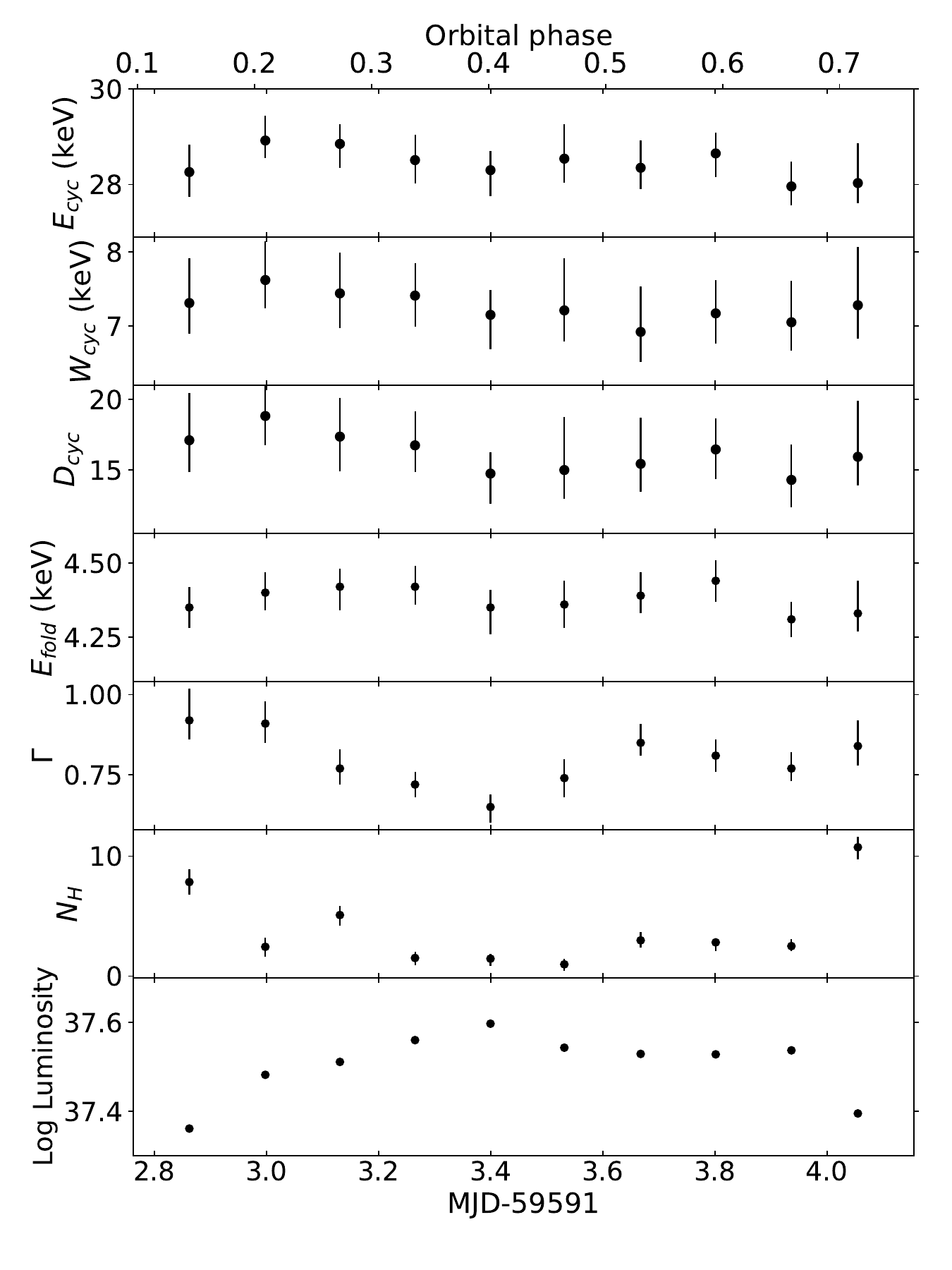} 
    \caption{{NPEX fits} showing the dependence of the CRSFs, the fold energy $E_{\rm fold}$, the photon index $\Gamma$, and the absorption column density $N_{\rm H}$ on the orbital phases for the first binary orbit and second orbit, respectively. The orbital phases are calculated from the orbital parameters provided by Fermi/GBM.}
    \label{fig:GE_orbit}
\end{figure*}

In this paper, we present a detailed timing and spectral analysis of Cen X-3 with observations 
made by NuSATR of two binary orbits covering two intensity states. Here, we focus on the dependence of the continuum and CRSF parameters on the luminosity and spectral variability during the orbital phases.

\subsection{Dependence of spectral parameters on luminosity}

We performed fitting on all 23 of the time intervals {with the FDCUT, HIGHECUT, and NPEX continuum models.} From all the fitting results, we {first} investigated the dependence of the continuum parameters  on luminosity in Cen X-3. The luminosity is calculated from the fitted unabsorbed flux of the spectra based on the source distance of $\sim$6.9 kpc \citep{2021MNRAS.507.3899V}. {As depicted in the left plots of Figure \cref{fig:cont_Flux}, for the results derived from the FDCUT model (the blue points), the photon index $\Gamma$, varying from 0.8 to 1.3, exhibits a positive correlation with luminosity; the Pearson correlation coefficient is about 0.86 with a probability (the probability of the observed data given that the null hypothesis of no correlation is true) of 1.54e-07, indicating that the continuum spectrum tends to soften  at a high accretion rate. Beyond the luminosity of $\sim 3 \times 10^{37}$ erg s$^{-1}$, the photon index increases slowly and remains flat; such spectral variability can be naturally ascribed to the changes in radiation pressure. The clustered $\Gamma$ in the range >$\sim 3 \times 10^{37}$ erg s$^{-1}$ may indicate that the accretion regime was close to the critical luminosity. \cite{2022ApJ...939...67B} suggested that their new model can explain the flatter, harder spectrum with increasing luminosity in Her X-1.  Based on the amount of PdV work on the radiation field, these authors
propose that in high-luminosity (supercritical) sources, the enhanced compression can lead to a flatter continuum. Conversely, a softer spectrum may indicate a subcritical source. The cut-off energy ranges from 20 to 26 keV and is positively related to luminosity; it begins to decrease around $ 3 \times 10^{37}$ erg s$^{-1}$. Such a high-energy quasi-exponential cutoff can be explained by the theoretical work of \cite{2007ApJ...654..435B}, who successfully applied their model to reproduce the phase-averaged X-ray spectra of Cen X-3. The fold energy varies from 4 to 6 keV, which is a signature for the electron temperature in the case of nonsaturated Comptonization. The electron temperature of 4.77 keV estimated by \cite{2021A&A...656A.105T}, who successfully applied the accretion column model of \citep{2007ApJ...654..435B} to Cen X-3, is very close to the $E_{\rm fold}$. The fold energy shows a decreasing trend at the beginning, followed by an increase, with a sudden decrease around $3 \times 10^{37}$ erg s$^{-1}$. The hydrogen column density $N_H$ varies from 1 to 85 $\times$ 10$^{22}$ atoms cm$^{-2}$, exhibits a negative correlation with the luminosity, and then remains stable, which is consistent with the results of \cite{Tamba_2023}, suggesting that the variations in absorption may affect the luminosity. {The covering fraction decreases from $\sim$0.8
to $\sim$0.5 with large fluctuations.} These spectral continuum parameters of the FDCUT model were found to show differences near $3 \times 10^{37}$ erg s$^{-1}$. The best fit for the FDCUT model yields a reduced $\chi^2$ of around 1.1. Regarding the HIGHECUT model (the black points in Figure \cref{fig:cont_Flux}), the behaviors of the photon index $\Gamma$ and $N_H$ are comparable with those seen when applying the FDCUT model. However, the fitting quality of HIGHECUT is worse at high luminosity and is not adequate to describe the data well.}

The right plots of Figure \cref{fig:cont_Flux} present the continuum parameters (the red points) for the NPEX model, which can also provide a good description of the data. Instead, the photon index varies between 1 and 3 and shows a negative correlation with luminosity, which is likely due to the evolution of E$_{\rm fold}$ and its effect on the continuum. The hydrogen column density decreases with increasing luminosity, consistent with the results obtained from FDCUT and HIGHECUT continuum models. At higher luminosities (>$3 \times 10^{37}$ erg s$^{-1}$), both $\Gamma$ and $N_H$ remain stable. Moreover, the folding energy also displays differences below and above this luminosity. These changes in spectral continuum parameters may indicate transitions between accretion regimes, such as from subcritical to critical. To explore this potential transition around $3 \times 10^{37}$ erg s$^{-1}$, we further investigated the dependency of CRSFs on luminosity.

The centroid energy of CRSFs in some X-ray pulsars has been observed to vary with X-ray luminosity, as discussed in a review by \citealt{2019A&A...622A..61S}. The discovery of a positive correlation between CRSF energy and luminosity at low and moderate luminosity was first made by \cite{2007A&A...465L..25S} when studying Her X-1. Similar trends have been identified in other sources, such as Vela X-1 \citep{2014ApJ...780..133F,2016MNRAS.463..185L,2022MNRAS.514.2805L}, Cep X-4 \citep{2017A&A...601A.126V}, and V0332+53 \citep{2017MNRAS.466.2143D,2018A&A...610A..88V}. Conversely, at higher luminosities, a distinct negative correlation was observed in V 0332+53 \citep{1990ApJ...365L..59M,2006MNRAS.371...19T}. Furthermore, 1A 0535+262 \citep{2021ApJ...917L..38K} has been confirmed to exhibit an anti-correlation above the critical luminosity and a positive correlation below the critical luminosity.

As shown in Figure \cref{fig:cyc_Flux}, {we present the CRSF parameters versus luminosity based on the FDCUT (the blue points) and NPEX (the red points) models. The width of the CRSF varies between 6 and 8 keV, and is correlated with the cyclotron line energy with a Pearson correlation coefficient of 0.69, and a probability of $\sim$ 0.0002 for the FDCUT model, and a Pearson correlation coefficient of 0.57 and a probability of $\sim$ 0.0043 for the NPEX model. The depth of the CRSF first shows an increasing trend and then decreases as the luminosity increases. There is probably a complex variation between X-ray luminosity and the CRSF energies for Cen X-3. For both continuum models, below $\sim 2 \times 10^{37}$ erg s$^{-1}$, there appears to be a decreasing trend between CRSF energy and luminosity, while above it, the CRSF energy begins to rise. However, at higher luminosities (>$\sim 3 \times 10^{37}$ erg s$^{-1}$),  a decreasing trend is seen. To further investigate the correlation between CRSF energy and luminosity around 3 $\times 10^{37}$ erg s$^{-1}$, we fitted the relationship of $E_{\rm cyc}- L_x$ of the FDCUT model in the range of $\sim2- 3 \times 10^{37}$ erg s$^{-1}$ and above the luminosity of 3 $\times 10^{37}$ erg s$^{-1}$ with a linear function separately. For the luminosity of 2-3 $\times 10^{37}$ erg s$^{-1}$, we fitted the relation with $E_{\rm cyc} = (1.33 \pm 0.34) \times (L_x/10^{37}) + (24.50 \pm 0.87)$ keV with a Pearson’s correlation coefficient r of 0.84 and a probability of around 0.0083, which shows a positive correlation. Above 3 $\times 10^{37}$ erg s$^{-1}$, we obtained the relation of $E_{\rm cyc} = (-0.69 \pm 0.24) \times (L_x/10^{37}) + (30.40 \pm 0.83)$ keV with a Pearson correlation coefficient of about -0.76 and a probability of about 0.028. For the NPEX model, the evolution of $E_{\rm cyc}- L_x$ is similar to the FDCUT model and gives comparable results, illustrating that the choice of continuum model does not significantly affect the relation between CRSF energy and luminosity. Therefore, the variation of $E_{\rm cyc}- L_x$ shows a possible positive and negative correlation below and above the transitional luminosity of $\sim 3 \times 10^{37}$ erg s$^{-1}$, respectively. However, we are unable to conclude that there is a distinct correlation or anticorrelation between the energy and luminosity of the CRSFs due to the large uncertainties of the centriod energy. \cite{2011A&A...535A.102M} reported that there is no transition in the accretion regime in Cen X-3. Recently, \cite{LIU2023} demonstrated that the CRSF energies have no correlation with luminosity in a larger range of $\sim (5-9) \times 10^{37}$ erg s$^{-1}$ based on Insight-HXMT observations obtained in 2022. \cite{2024MNRAS.527.6981D} claimed an anticorrelation between CRSF energy and flux in their work with the same NuSTAR data as that analyzed in the present study, while using a continuum with a "newhcut" cutoff. The differences between our findings and those of \cite{2024MNRAS.527.6981D} might stem from their choice of continuum and specifically their inclusion of a second absorption feature, for which we see no need in our analysis.} Therefore, the relationship between cyclotron line energy and luminosity is still not clear. In the following, we discuss the possible critical luminosity of Cen X-3 based on different theories.

A transition between subcritical and supercritical accretion regimes was clearly observed in the transient X-ray binaries V 0332+53 and 1A 0535+262 (see \citealt{2022arXiv220414185M}). As predicted by \cite{2012A&A...544A.123B} and \cite{2015MNRAS.447.1847M}, in the supercritical case ---that is, above a certain critical luminosity $L_{\rm crit}$---, higher accretion rates lead to higher radiation pressure, and the radiation-dominated shock decelerates the falling material, forming the accretion column \citep{1976MNRAS.175..395B}, and the emission height in the column increases with luminosity. Below the critical luminosity (subcritical), the Coulomb interaction dominates the deceleration, and the height of the line-forming region reduces with luminosity. A change of height is associated with a variation in the local magnetic field strength. Therefore, at high luminosity, the CRSF energy is expected to decrease with the luminosity. Nevertheless, at low accretion rates, a positive correlation is expected. This positive correlation below the critical luminosity may also result from the Doppler effect proposed by \cite{2015MNRAS.454.2714M}. The transition from the subcritical to the critical regime may be related to the change in the emission geometry{} from a mixture of pencil and fan beams to only fan beams, consistent with change in the shape of the pulse profiles at low states to high states in Figure \cref{fig:pulse_profile}.  \citet{2012A&A...544A.123B}, based on their theoretical work, show that critical luminosity can be expressed as 
\begin{equation}
\begin{aligned}
L_{\text {crit}}\simeq 1.49 \times 10^{37} \ \mathrm{erg} \mathrm{s}^{-1}\left(\frac{\Lambda}{0.1}\right)^{-7 / 5} \times B_{12}^{16 / 15},
\end{aligned}
\end{equation}
for ${M_{\rm NS}}$ =$1.4 M_{\odot}$ and ${R_{\rm NS}}$ =10 km. {We assume $\Lambda$ = 0.1 for the case of disk accretion, and compute the surface magnetic field strength $B_{12}$  using the fundamental energy as the surface value for CRSF energy. We estimate the expected $L_{\text {crit}}$ to be equal to $\sim 4 \times 10^{37} \ \mathrm{erg} \mathrm{s}^{-1}$, which is approximately the observed transitional luminosity of 3 $\times 10^{37}$ erg s$^{-1}$. {In our calculations of $L_{\text {crit}}$, we neglect the effect of gravitational redshift (e.g., a factor of $(1+z)^{-2}$), which introduces additional uncertainties in the theoretical estimates when comparing to observations.} If we adopt the distance of 8.0 kpc \citep{1974ApJ...192L.135K}, the observed transitional luminosity can be up to 4 $\times 10^{37}$ erg s$^{-1}$, which is also consistent with the expected $L_{\text {crit}}$. According to the theoretical value estimated by \cite{2015MNRAS.447.1847M}, the critical luminosity is around $10^{37}$ erg s$^{-1}$ for $l_0/l =$ 1.0 and the pure X-mode polarization, which is a little smaller than the value derived by \cite{2012A&A...544A.123B}. Although the expected negative correlation between the CRSF energy and the luminosity above the $10^{37}$ erg s$^{-1}$ is not seen, these theoretical results provide a good description of the accretion process, and more high-quality data are needed to verify these models in the future. Based on the theoretical critical luminosity and also considering the different behaviors of the shapes of the pulse profiles and hardness ratios from the low rate to the high rate, it is possible that the critical luminosity is contained within the range of $\sim (0.5-4)\times 10^{37}$ erg s$^{-1}$.}


\subsection{Spectral variability over orbital phases}

Cen X-3 has a relatively short orbital period, and the spectral parameters probably depend on the orbital phases. Here we present spectral parameters over the two binary orbits observed in 2022 based on the {NPEX model} fitting results in Section \cref{sec:result}, which cover two intensity states. Figure \cref{fig:GE_orbit} shows the results of the evolution of CRSF energies and continuum parameters over the two orbits.

Over the first binary orbit ($\sim (1.0-2.5)\times 10^{37}$ erg s$^{-1}$, the left plots of Figure \cref{fig:GE_orbit}), the CRSF energy varies between 27 keV and 29 keV. The evolution of the width and depth of the CSRFs is similar to the variation in their energy. The fold energy $E_{\rm fold}$ also roughly follows the CRSF energy. In addition, the behavior of the photon index $\Gamma$ from 1.0 to 2.0 is opposite to luminosity. The photoelectric absorption $N_{\rm H}$ shows a large variation between 10 $\times$ 10$^{22}$ atoms cm$^{-2}$ and 20 $\times$ 10$^{22}$ atoms cm$^{-2}$ over the orbit. We also note that $N_{\rm H}$ slightly decreases near the beginning of the orbit and increases after that. The increase in $N_{\rm H}$ near an eclipse may be due to a larger amount of stellar material within the line of sight \citep{1988ApJ...324..974C}. An increase in $N_{\rm H}$ around the phase of 0.3-0.4 is observed, which is probably due to the bow shock, as mentioned by \cite{2008ApJ...675.1487S}.

Over the second orbit ($\sim (2.5-4)\times 10^{37}$ erg s$^{-1}$; see the right plots of Figure \cref{fig:GE_orbit}), the variations in CRSF energies of 28-29 keV are almost opposite to the evolution of luminosity, except for the two points in luminosity of <3 $\times 10^{37}$ erg s$^{-1}$. Furthermore, the widths and depths of the CRSFs, and the fold energy $E_{\rm fold}$ , roughly follow the shape of the energy of the CRSFs. Here, the $\Gamma$ from 0.1 to 1.0 has a smaller value compared with the first orbit and also shows an anticorrelation with luminosity. The variability of the absorption column $N_{\rm H}$ around the value of 1-10 $\times$ 10$^{22}$ atoms cm$^{-2}$ is similar to the first orbit and also shows a small increase around phase 0.3. 

Since the highly variable $N_{\rm H}$ probably originates from absorption by the stellar wind, we considered the possible stellar wind structure other than the accretion stream. When the majority of the material in the wind passes through our line of sight, variations in absorption are expected. In general,  $N_{\rm H}$ increases after the orbital midpoint, as predicted by the simple wind model, which is consistent with our results. For example, enhanced absorption is seen in Vela X-1 \citep{1990ApJ...361..225H,2022A&A...660A..19D}, which shows a strong stellar wind and tidal stream \citep{1991ApJ...371..684B}. 


\section{Conclusion} \label{sec:summary}
We carried out a detailed temporal and spectral analysis of the accreting X-ray pulsar Cen X-3 using NuSTAR observations obtained in 2022. The pulse profiles exhibit remarkable stability throughout the orbital phase. These profiles demonstrate energy dependence, displaying double peaks in the energy band below 15 keV and a single peak in the higher energy band, with the pulse fraction increasing with energy. Notably, the profile in the energy band of 3-5 keV during a high-intensity state exhibited marginal triple peaks, with the main peak further divided into two subpeaks. Additionally, we highlight a clear positive correlation between pulse fraction and flux.

From our spectral analysis, we identify a prominent cyclotron line with centroid energies ranging from 26 keV to 29 keV using different continuum models, and the spectra in Cen X-3 are well described by the FDCUT and NPEX models. Additionally, we studied the dependence of the cyclotron line energies at different luminosities and discuss the theoretical critical luminosity based on the different models. Regarding the orbital-phase evolution of the spectra, the spectral parameters display variability throughout the orbital phase, and the highly variable photoelectric absorption may suggest the presence of clumpy stellar winds.

\begin{acknowledgements}
We are grateful to the referee for the useful suggestions to improve the manuscript. This work is supported by the NSFC (12133007) and the National Key Research and Development Program of China (No. 2021YFA0718503). This work has made use of archival data provided by the High Energy Astrophysics Science Archive Research Center (HEASARC).
\end{acknowledgements}


\bibliographystyle{aa}
\bibliography{refer}

\end{document}